\documentclass[journal]{IEEEtran}
\usepackage[utf8]{inputenc}
\usepackage{echodefs}
\usepackage{booktabs}
\usepackage[caption=false]{subfig}
\usepackage{import}

\definecolor{maroon}{RGB}{181, 23, 0}
\definecolor{easy_blue}{RGB}{0, 118, 186}

\newcommand{\rev}[1]{{\color{black}#1}}
\newcommand{\aqe}[1]{{\color{black}#1}}

\title{Differentiable Uncalibrated Imaging}

\author{
Sidharth Gupta$^{\,\dagger}$, Konik Kothari$^{\,\dagger}$, Valentin Debarnot$^{\,\ddagger}$, and Ivan Dokmani\'c$^{\,\ddagger,\, \dagger}$ \\
\normalsize{$^\dagger\,$University of Illinois at Urbana-Champaign, $^\ddagger\,$University of Basel} 
\\
{\texttt{\{gupta67, kkothar3\}@illinois.edu, \{valentin.debarnot, ivan.dokmanic\}@unibas.ch}}
\thanks{This work was supported by the European Research Council Starting Grant 852821---SWING.
This research was also supported by the donation of GPUs from NVIDIA Corporation to the University of Illinois at Urbana-Champaign.
Additionally, calculations were performed at sciCORE (\url{http://scicore.unibas.ch/}) scientific computing center at University of Basel.
}
}

\begin{document}

\maketitle  

\begin{abstract}

We propose a differentiable imaging framework to address uncertainty in \textit{measurement coordinates} such as sensor locations and projection angles. We formulate the problem as  measurement interpolation at unknown nodes supervised through the forward operator. 
To solve it we apply implicit neural networks, also known as neural fields, which are naturally differentiable with respect to the input coordinates. We also develop differentiable spline interpolators which perform as well as neural networks, require less time to optimize and have well-understood properties. Differentiability is key as it allows us to jointly fit a measurement representation, optimize over the uncertain measurement coordinates, and perform image reconstruction which in turn ensures consistent calibration. We apply our approach to 2D and 3D computed tomography, and show that it produces improved reconstructions compared to baselines that do not account for the lack of calibration. The flexibility of the proposed framework makes it easy to \rev{extend} to almost arbitrary imaging problems.

\end{abstract}

\section{Introduction} \label{sec:introduction}

In computational imaging a physical process, $\mathcal{A}$, such as 2D computed tomography (CT) relates the object we want to image, $x$, to an observable field, $y$. Both $x$ and $y$ are naturally functions of continuous coordinates: $x$ could be a density over 2D spatial coordinates (e.g., $[0, 1]^2$), and $y$ a sinogram over angles and 1D projection coordinates (e.g., $[0, \pi) \times [-1, 1])$.

In practice, however, we have finite sensors. Denoting the space of continuous measurement coordinates by $\Omega$, the sensors sample the field $y$ at $\wt{\vmu} = (\wt{\mu}_1, \ldots, \wt{\mu}_M) \in \Omega^M$, such that the observed measurements $\vy \in \R^M$ follow
\begin{align}
    \vy = \mA_{\wt{\vmu}}(x) + \veta, \label{eq:problem}
\end{align}
where
\begin{align}
    \vy \bydef \vy(\wt{\vmu}) \bydef [y(\wt{\mu}_1), \ldots, y(\wt{\mu}_M)]^\T,
\end{align}
and $\veta \in \R^M$ is the measurement noise.
The discrete forward operator $\mA_{\wt{\vmu}}$ parameterized by $\wt{\vmu}$ samples the output of $\mathcal{A}$. In computational imaging we work with a discretization or some other finite-dimensional approximation of $x$ denoted by $\vx \in \R^N$. Hereafter we let $\mA_{\wt{\vmu}}$ act on $\vx$ rather than on $x$. 

In this paper we address the scenario where the true measurement coordinates $\wt{\vmu}$ are only approximately known: the imaging system is out of calibration. Consequently, the true operator $\mA_{\wt{\vmu}}$ is unknown, and we work with an operator $\mA_{\vmu}$ for measurement coordinates $\vmu = (\mu_1, \ldots, \mu_M)$, which \textit{would} be correct if the system was calibrated. The assumed measurement coordinates $\vmu$ are related to the \emph{unknown} true measurement coordinates $\wt{\vmu}$ by small perturbations. 

Not accounting for the mismatch between $\wt{\vmu}$ and $\vmu$ can lead to a poor reconstruction. The gist of our method is to learn a representation of the measurement space that can be evaluated and differentiated at arbitrary measurement coordinates $\vmu$. This gives us measurements $\vy({\vmu})$ that are then well-suited for a reconstruction method that uses $\vmu$. Our proposed framework enables us to
\begin{enumerate}
    \item Jointly reconstruct the image $\vx$ and learn the true unknown measurement coordinates $\wt{\vmu}$ by using gradient-based optimization.
    \item Learn continuous measurement representations that take measurement coordinates as input. These representations can be evaluated at $\vmu$ and are in essence interpolations of the discrete measurements with unknown interpolation knots.
    \item Leverage standard differentiable image reconstruction methods that exist for the assumed $\vmu$ even though the observations correspond to unknown $\wt{\vmu}$. This helps learn consistent measurement representations.
\end{enumerate}

A strength of our approach is that we can use any continuous interpolation method that admits backpropagation to input coordinates. To showcase this, we use implicit neural networks (neural fields) but also develop fully-differentiable variable-knot splines which allow us to optimize spline control points and control point weights. We show that splines perform as well as implicit neural networks while being faster to fit and simpler to interpret. A key property of both these representation types is that we can perform automatic differentiation with respect to their input coordinates and recover $\wt{\vmu}$ by gradient-based optimization.

\subsection{Example: Computed tomography} \label{sec:ct_example}

\begin{figure*}[t]
    \centering
    \def\svgwidth{0.95\linewidth}
    \fontsize{8.5}{8.5}\selectfont
    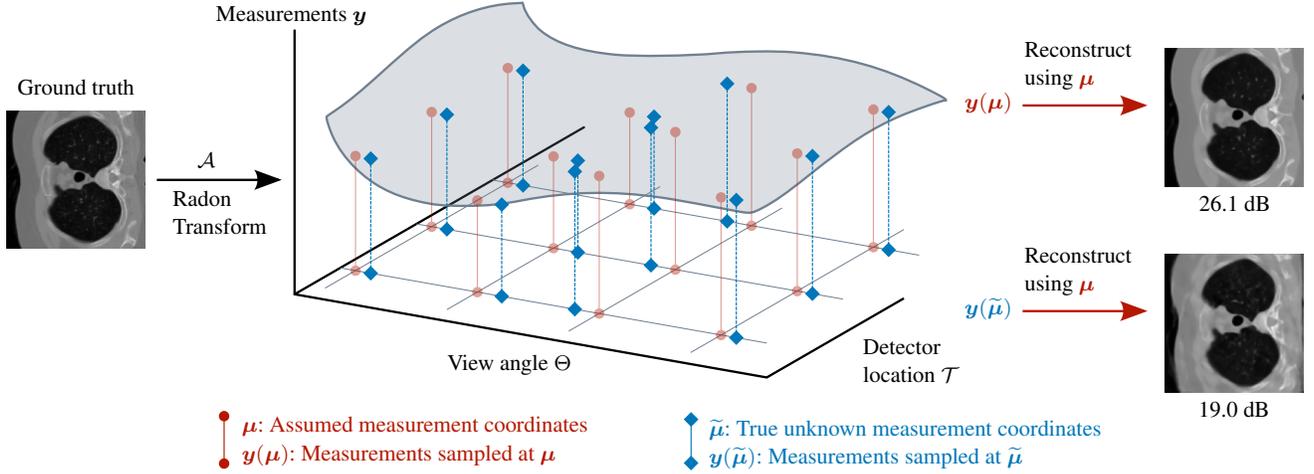
    \caption{Visualization of the parameter mismatch between $\wt{\vmu}$ and $\vmu$ in 2D CT imaging \rev{that is described in Section \ref{sec:ct_example}}. A mismatch in the view angles between $\vmu$ and $\tilde\vmu$ can produce a significant drop in reconstruction quality. }
    \label{fig:mismatch_illustration}
\end{figure*}

We illustrate our framework with 2D CT, where we measure parallel beam projections of an image at different view angles in a detector plane. Let $\Omega = [0, \pi) \times \R$, $x \in L^2([-1,1]^2)$, and $y \in L^2(\Omega)$.\footnote{We denote by $L^2(\calM)$ the space of square integrable functions on $\calM$.} The measured projections $\vy$ are obtained from a finite set of discrete view angles $\wt{\Theta} = (\wt{\theta}_1, \ldots, \wt{\theta}_J)$, and are sampled at a finite set of discrete detector locations $\wt{\calT} = (\wt{t}_1, \ldots, \wt{t}_K)$. The true set of unknown measurement coordinates are then $(\wt{\mu}_m)_{m=1}^M = \wt{\Theta} \times \wt{\calT}$ with $M = J \cdot K$. Furthermore, $\mA_{\wt{\vmu}}$ is given by the discrete Radon transform for view angles $\wt{\Theta}$ and detector locations $\wt{\calT}$. Many other computational imaging applications such as electron cryotomography (CryoET), magnetic resonance imaging and optical microscopy can be parameterized similarly.

\begin{figure}[t]
    \centering
    \includegraphics[width=1\linewidth]{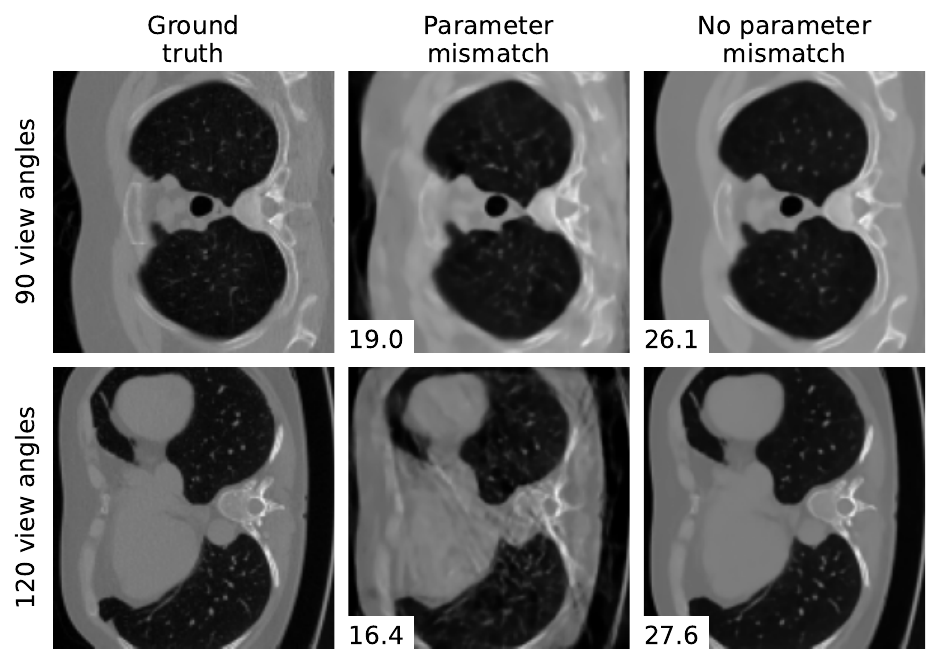}
    \caption{2D CT reconstructions with 90 and 120 view angle. The SNRs (in dB) are written in the bottom-left corners. When there is parameter mismatch because the true and assumed view angles differ, the reconstruction method performs poorly and there is a clear degradation.}
    \label{fig:parameter_mismatch}
\end{figure}

The parameter mismatch between $\wt{\vmu}$ and $\vmu$ and its severe impact is illustrated in Fig. \ref{fig:mismatch_illustration}.
We use a state-of-the-art reconstruction method that computes a filtered backprojection estimate with $\vmu$ and feeds it into a UNet deep neural network to obtain a reconstruction of $\vx$ \cite{ronneberger2015u, jin2017deep}. The neural network is trained in a supervised manner using training data that is also generated using $\vmu$.
When there is parameter mismatch, the true view angles differ from the assumed view angles and therefore $\wt{\vmu} \neq \vmu$. Fig. \ref{fig:mismatch_illustration} shows that in this case, the reconstruction method produces a degraded reconstruction from both a visual and SNR evaluation. Fig. \ref{fig:parameter_mismatch} shows this for another ground truth image and number of view angles. The last column of Fig. \ref{fig:parameter_mismatch} also shows that when there is no parameter mismatch ($\wt{\vmu} = \vmu$), the same reconstruction method performs strongly.

\subsection{Related work}

Many of the state-of-the-art methods for solving imaging inverse problems are based on deep learning \cite{mccann2017convolutional, ongie2020deep}. Popular supervised approaches use the forward operator to obtain an initial estimate which is then enhanced by a neural network 
(reconstruction method in Fig. \ref{fig:parameter_mismatch} for example)
\cite{jin2017deep, antholzer2019deep, kothari2018random, ulyanov2018deep}. Unrolling iterative methods or replacing optimization components with deep networks is another established approach
\cite{gregor2010learning, adler2018learned, gupta2018cnn, rick2017one, gilton2019neumann, zhang2021plug}.
A different direction involves requiring the inverse problem solution to lie in the range of a generative neural network \cite{bora2017compressed,kothari2021trumpets,song2021solving}. In general all these methods utilize a forward operator in some way but do not account for measurement coordinate uncertainty.
This can severely impact the reconstruction as shown in Fig. \ref{fig:parameter_mismatch}. 
In this paper we propose to address this problem by reevaluating the measurements at the measurement coordinates that deep neural network reconstruction methods are designed for.

While much less common, there are some recent deep learning approaches which address measurement coordinate uncertainty. Gilton et al. explored fine-tuning a neural network that was trained with measurements sampled at $\vmu$ to work well with test measurements sampled at unknown $\wt{\vmu}$ \cite{gilton2021model}. Our method is different: we reevaluate the measurements via a measurement representation for a single example so we do not need to fine-tune. Another appealing approach is to train a neural network on a family of operators with different parameterizations \cite{gossard2022training}. These methods always induce a tradeoff between the reconstruction quality and the variety of forward maps they are trained on, and they only work well for the distribution of perturbations seen at training time. There is also the challenge of dataset generation, especially in compute-intensive problems. We mitigate this tradeoff by using consistency to identify the true measurement coordinates.

Continuous representations have been used to solve a variety of science and engineering problems. To the best of our knowledge no prior work has used them for imaging with measurement coordinate uncertainty. Splines have long been used to model surfaces and curves in geometry \cite{bartels1995introduction, rogers1989mathematical, piegl1991nurbs, cohen1980discrete}. Moreover, differentiable splines with parameters that can be fitted using automatic differentiation have also been developed \cite{prasad2022nurbs}. Recently deep learning methods called neural fields or implicit neural networks have proven to be extremely good and efficient continuous representations \cite{xie2022neural}. They have been employed in a broad spectrum of applications ranging from representing geometry via signed distance functions \cite{park2019deepsdf, mildenhall2020nerf} to solving partial differential equations \cite{sitzmann2020implicit}. Implicit neural networks have also been used to solve tomographic imaging inverse problems \cite{Reed_2021_ICCV,lozenski2022neural}. In particular, Sun et al. also used them to represent 2D CT measurements \cite{sun2021coil}. Rather than for calibration, they use implicit networks to upsample measurements for downstream reconstruction by a deep neural network. While upsampling may also be performed by traditional tools such as splines, our framework relies on the differentiability of the used representations---both neural and spline-based---with respect to the input coordinates.

\rev{In this paper, we learn forward model parameters and continuous measurement representations through a differentiable imaging framework \cite{chen2023differentiable}. We use automatic differentiation to optimize the parameters and inputs of implicit neural network and differentiable spline representations. Prior differentiable imaging works for microscopy \cite{du2020three}, holography, ptychography and ptychographic tomography \cite{du2021adorym} also learn forward model parameters via automatic differentiation. However, these works do not learn a continuous measurement representation, and so do not reevaluate measurements for use with state-of-the-art reconstruction methods that assume different parameters.}

\rev{There are alternative approaches for handling uncertainty in measurement coordinates. Bundle adjustment is an example from computer vision where scene coordinates, camera coordinates and system coordinates are jointly optimized \cite{triggs2000bundle}. Similar to our framework, bundle adjustment uses a good starting estimate to optimize the coordinates. However, a key difference is that we obtain a final solution by reevaluating measurements at the assumed measurement coordinates, and use a reconstruction method designed for the assumed measurement coordinates. Imaging approaches based on alternating minimization optimization \cite{campisi2017blind}, and designing handcrafted regularizers to manage ill-posedness \cite{chan1998total,riis2021computed} have also been employed when there is miscalibration. In this work we jointly minimize our objective function. Furthermore, we use a consistency loss which can be interpreted as a regularizer when learning the measurement representation.}

Our work is related to the broader theme of inverse problems with ``noisy'' forward operators. On one end of the spectrum there are inverse problems where the operator (and $\wt{\vmu}$) is perfectly known, and at the other extreme there are blind inverse problems where the operator and $\wt{\vmu}$ are completely unknown. Related blind imaging problems are tomography with unknown view angles \cite{basu2000uniqueness, coifman2008graph} and cryo-electron microscopy with unknown projection angles \cite{bendory2020single}. In between the extremes, there are semi-blind inverse problems where the operator and $\wt{\vmu}$ are approximately known. Total least squares approaches that perturb an assumed operator \cite{golub1980analysis, markovsky2007overview, gupta2021total} and our proposed measurement coordinate-based framework fall under this category. 

\subsection{Paper organization}

In Section \ref{sec:framework} we present the optimization problem which models joint calibration and image reconstruction. We first model measurements as continuous functions and then show how to learn these measurement representations. We show how implicit neural networks and splines can both be used as representations. Section \ref{sec:experiments} \rev{numerically} verifies that when there is measurement coordinate uncertainty, our proposed method yields significantly improved reconstructions. The \rev{simulated} experiments are performed on \rev{2D CT in the main paper and 3D CT in Appendix \ref{app:ct3d_experiments}}. We conclude the paper and motivate future work directions in Section \ref{sec:conclusion}.

\section{Differentiable framework} \label{sec:framework}

We wish to learn a continuous representation of the  measurement space so that we can sample it at specific measurement coordinates and obtain the corresponding measurements. We model the measurement representations as continuous functions $r_{\vvarphi}(\cdot)$ that take measurement coordinates as input and map
them to the corresponding sampled measurements. The learnable parameters, $\vvarphi \in \Phi$ where $\Phi$ is the space of feasible parameters, are optimized so that
\begin{align}
    r_{\vvarphi}(\omega) \approx \vy(\omega) \label{eq:measurement_reps}
\end{align}
where $\omega \in \Omega$ is a measurement coordinate and $\vy(\omega) \in \R$ is the sample from the measurement space at measurement coordinate $\omega$. For convenience we also denote a batch evaluation of $r_{\vvarphi}(\cdot)$ as
\begin{align}
    R_{\vvarphi}(\vomega) & \bydef [r_{\vvarphi}(\omega_1), \ldots, r_{\vvarphi}(\omega_Q)]^\T \notag \\
    & \approx [\vy(\omega_1), \ldots, \vy(\omega_Q)]^\T
\end{align}
where $\vomega = (\omega_1, \ldots, \omega_Q) \in \Omega^Q$ and $R_{\vvarphi}(\vomega) \in \R^Q$. We also require $r_{\vvarphi}(\cdot)$ to be differentiable with respect to $\vvarphi$ and its input so that we can use gradient-based optimization to estimate $\vvarphi$ and the unknown measurement coordinates.

\subsection{Joint optimization objective}

We want $r_{\vvarphi}(\cdot)$ to accurately produce samples from the space of measurements. Since we only observe measurements $\vy$ at measurement coordinates $\wt{\vmu}$ in \eqref{eq:problem}, we require 
\begin{align}
    R_{\vvarphi}(\wt{\vmu}) \approx \vy
\end{align}
However, since $\wt{\vmu}$ is unknown we cannot use it to verify the accuracy of $r_{\vvarphi}(\cdot)$. This motivates us to jointly learn the representation parameters $\vvarphi$ and the unknown measurement coordinates $\wt{\vmu}$ by minimizing a measurement fitting loss,
\begin{align}
    \calL_{\text{fitting}}(\vnu, \vvarphi) \bydef \norm{\vy - R_{\vvarphi}(\vnu)}^2_2 , \label{eq:loss_fitting}
\end{align}
with respect to the input $\vnu \in \Omega^M$ and $\vvarphi$. Recall $\vy \in \R^M$ is defined by~\eqref{eq:problem}.

Learning $r_{\vvarphi}(\cdot)$ by minimizing only $\calL_{\text{fitting}}(\vnu, \vvarphi)$ with respect to both the representation's parameters and input would not in general result in an accurate measurement representation because there are too many degrees of freedom. Therefore we regularize and control $\vvarphi$ by enforcing $r_{\vvarphi}(\cdot)$ to be consistent with reconstructions that could be obtained by using its output. This is done by minimizing a consistency loss with respect to~$\vvarphi$,
\begin{align}
    \calL_{\text{consistency}}(\vvarphi) \bydef \norm{R_{\vvarphi}(\vmu) - \mA_{\vmu} G_{\vmu}(R_{\vvarphi}(\vmu))}^2_2 , \label{eq:loss_consistency}
\end{align}
where $G_{\vmu}: \R^M \rightarrow \R^N$ is a differentiable reconstruction method that was designed using measurement coordinates $\vmu$. Putting everything together, our complete joint optimization objective is
\rev{
\begin{align}
    \wh{\vmu},\, \wh{\vvarphi} = \argmin_{\vnu \in \Omega^M, \, \vvarphi \in \Phi} 
     \, \calL_{\text{fitting}}(\vnu, \vvarphi) 
    + 
    \lambda \, \calL_{\text{consistency}}(\vvarphi) \label{eq:joint_objective}
\end{align}
where $\lambda \in \mathbb{R}_{\geq 0}$ is a tunable weight that controls the relative importance of the consistency and fitting losses.}
As the assumed coordinates $\vmu$ are close to the true unknown coordinates $\wt{\vmu}$, we initialize $\vnu$ to $\vmu$. After completing the optimization \eqref{eq:joint_objective}, the learned coordinates $\wh{\vmu}$ are close to $\wt{\vmu}$ (verified in Section \ref{sec:experiments}), and the final reconstruction and estimate of $\vx \in \R^N$ is given by
\begin{align}
    \wh{\vx} = G_{\vmu}(R_{\vvarphi}(\vmu)) . \label{eq:solution}
\end{align}

\rev{One of the advantages of the objective \eqref{eq:joint_objective} is that the same operator, $\mA_{\vmu}$ in \eqref{eq:loss_consistency}, is used in each optimization iteration. A formulation that uses the continually updating learned measurement coordinates $\vnu$ in \eqref{eq:loss_consistency}, would require the operator to be rebuilt after each optimization iteration. \aqe{If there is no efficient implementation to change the measurement coordinates due to the complexity of the forward process, and if the operator is large, this can be severely time consuming and resource intensive.
}
}

\subsection{Leveraging reconstruction methods}

A key aspect of our framework is that consistency and the final reconstruction are obtained by using reconstruction method $G_{\vmu}(\cdot)$ that was designed using measurement coordinates $\vmu$ even though the observations in \eqref{eq:problem} are sampled at measurement coordinates $\wt{\vmu}$. This provides significant flexibility and allows us to incorporate a variety of reconstruction methods. For example, the reconstruction method may be a relatively straightforward adjoint or pseudoinverse operation. Alternatively, it can be a more complex neural network that provides state-of-the-art reconstructions with measurements from $\vmu$ (reconstruction method in Fig. \ref{fig:parameter_mismatch} is one example). Our framework is particularly advantageous in this case because it may be cumbersome to retrain a neural network for different measurement coordinates. 

\subsection{Measurement representations}

In order to better understand how to use the framework to solve real imaging problems, we now pick implicit neural networks and splines and explain how they are suitable measurement representations. 
While we consider these, we emphasize that our framework is general and is not restricted to these representation types.

\subsubsection{Implicit neural representations}

Implicit neural representations are deep feedforward neural networks that represent discrete signals as continuous functions. When used as the measurement representation $r_{\vvarphi}(\cdot)$ in \eqref{eq:measurement_reps}, $\vvarphi$ are the trainable network parameters. The input to the network are measurement coordinates and the output are the corresponding measurements. Implicit neural networks have previously been used to represent measurements when there is no measurement uncertainty \cite{sun2021coil}. In this case the network input was not optimized and consistency \eqref{eq:loss_consistency} was not enforced.

As implicit neural representations are neural networks, we can use automatic differentiation to calculate their gradients with respect to $\vvarphi$ and their input coordinates to solve \eqref{eq:joint_objective}. In this paper we use an architecture comprising a Fourier feature mapping layer followed by standard fully-connected layers \cite{mildenhall2020nerf, sun2021coil}.

\subsubsection{Differentiable splines} \label{sec:splines}

Splines use locally supported basis functions to represent signals as a continuous surface. In this paper we focus on Non-uniform Rational Basis Splines (NURBS) because of their ability to model complex surfaces~\cite{piegl1991nurbs, piegl1996nurbs, rogers2001introduction}. To aid understanding, we explain NURBS using the 2D CT imaging example that was introduced in Section~\ref{sec:ct_example}. The measurement coordinates $\vmu$ are the Cartesian product of assumed view angles and assumed detector locations, $\Theta \times \calT$ where $\Theta = (\theta_1, \ldots, \theta_J)$ and $\calT = (t_1, \ldots, t_K)$. The NURBS surface $s_{\vvarphi}(\cdot)$ with parameters $\vvarphi$ evaluated at measurement coordinate $\mu' = [\theta', t']^\T \in [0, \pi] \times \R$ is then
\begin{align}
    s_{\vvarphi}(\mu') = 
    \sum_{j=0}^{J - 1}
    \sum_{k=0}^{K - 1} 
    b_{j,k}(\mu') 
    \,
    \vp_{j,k} \label{eq:nurbs_basic}
\end{align}
where $b_{j,k}(\cdot) \in \R$ are scalar-valued rational basis functions with local support and $\vp_{j,k} \in \R^3$ are control point vectors. The NURBS are parameterized by weight parameters $w_{j,k} \in \R$ of the rational basis functions, and the control point vectors. This gives $\vvarphi = \set{w_{j,k}}_{j=0, \, k=0}^{J-1, \, K-1} \cup \{\vp_{j,k}\}_{j=0, \, k=0}^{J-1, \, K-1}$. Note that the control point vectors are three-dimensional because for each two-dimensional measurement coordinate $\mu'$, there is a corresponding scalar measurement. Consequently, according to~\eqref{eq:nurbs_basic}, $s_{\vvarphi}(\mu')$ is also three-dimensional. We then establish the following relationship between NURBS surfaces and our measurement representations \eqref{eq:measurement_reps} to get a spline measurement representation,
\begin{align}
    r_{\vvarphi}(\omega) = s_{\vvarphi}^\diamond(\omega) , \label{eq:rep_spline}
\end{align}
where $s_{\vvarphi}^\diamond(\cdot)$ denotes the value of the measurement dimension of $s_{\vvarphi}(\cdot)$.

It has recently been shown that automatic differentiation can be used to learn spline parameters \cite{prasad2022nurbs}. Hence, we develop differentiable spline representations and use gradient-based optimization to learn the NURBS parameters $\vvarphi$ and their input measurement coordinates. These differentiable splines fit into our framework straightforwardly as measurement representations \eqref{eq:rep_spline}, and we can use them to solve \eqref{eq:joint_objective}.

In our \rev{numerical} experiments, we carefully initialize the spline parameters $\vvarphi$ and then learn them: $w_{j,k}$ is initialized to one and control point vectors $\vp_{j,k}$ are initialized using the assumed measurement coordinates and their corresponding observed measurements, $\vp_{j,k} = \left[\theta_j, t_k, \vy([\wt{\theta}_j, \wt{t}_k]^\T)\right]^\T \in \R^3$. Additionally, in our \rev{simulated} experiments we also extend \eqref{eq:nurbs_basic} to higher dimensional measurement coordinates (see Appendix~\ref{app:ct3d_experiments}). Further details on NURBS and their implementation are provided in Appendices \ref{app:splines} and \ref{app:parameters}.

\section{Numerical verification with 2D CT imaging}
\label{sec:experiments}

\rev{We experimentally verify our framework by solving 2D CT imaging problems in the main paper. Further simulation results for 3D CT imaging are provided in Appendix~\ref{app:ct3d_experiments}.} 
\aqe{The imaging forward operators are implemented using the Operator Discretization Library (ODL) \cite{adler2017operator}.}
These \rev{simulated} experiments demonstrate that our framework can be used to solve imaging problems with different measurement coordinate dimensions. Furthermore, we exhibit the flexibility of our method by using implicit neural networks and splines as measurement representations.\footnote{Code available at \url{https://github.com/swing-research/differentiable_uncalibrated_imaging}.}

We use SNR (\rev{in} dB) to quantify the measurement noise and measurement coordinate uncertainty. SNR is calculated by
\begin{align}
    \text{SNR}(\vc,\, \vd) = -20 \log_{10} \left(\frac{\norm{\vc - \vd}_2}{\norm{\vd}_2} \right) .
\end{align}
If we let $\vy \in \R^M$ denote the observed measurements as in \eqref{eq:problem} and let $\wt{\vy}$ denote the unobserved noiseless measurements, the measurement noise level is $\text{SNR}(\vy,\, \wt{\vy})$. The measurement noise is simulated with zero-mean iid Gaussian noise with variance adjusted to achieve a target SNR level.

Due to their state-of-the art performance when there is no measurement coordinate uncertainty, we use deep neural networks for the reconstruction method, $G_{\vmu}(\cdot)$, in \eqref{eq:joint_objective}. For 2D CT we use a 2D Unet \cite{jin2017deep}, and for 3D CT \rev{in Appendix \ref{app:ct3d_experiments},} we use a 3D Unet \cite{cciccek20163d}. Following standard practice, a preprocessing step applies the pseudoinverse of the imaging operator to the measurements to produce an initial image estimate \cite{jin2017deep}. These networks are then trained in a supervised manner to map the initial estimates to ground truth images. The training data generation and preprocessing step are done with the assumed measurement coordinates $\vmu$. The measurements in the training data are noisy and in each experiment we use a Unet whose training measurement noise level matches the measurement noise level of the obtained measurements.

As mentioned in Section \ref{sec:framework}, the solution, $\wh{\vx}$, is given by~\eqref{eq:solution}. We use SNR to evaluate the solution quality, $\text{SNR}(\wh{\vx},\, \vx)$. We compare $\wh{\vx}$ against baseline reconstructions that are obtained by directly using the obtained measurements with the reconstruction Unets,
\begin{align}
    \vx_{\text{baseline}} = G_{\vmu}(\vy) .
\end{align}
Recall, the obtained measurements correspond to measurement coordinates~$\wt{\vmu}$. Appendix \ref{app:parameters} contains further hyperparameter and implementation details. 

In 2D CT imaging, one-dimensional projections of a two-dimensional object at different view angles are collected \rev{by an array of detectors. Our goal is to reconstruct an image of the object from these projections when there is uncertainty in only the view angles, only the detector locations, or in both the view angles and the detector locations at the same time.}

\rev{For these simulated experiments, the assumed view angles are uniformly spaced on the interval $[0, \pi]$, and the true view angles can have an unknown perturbation from these assumed view angles to simulate experimental incertitude. We keep the view angle measurement coordinate uncertainty level the same for each view angle---each true view angle is independently perturbed from the assumed view angle by zero-mean Gaussian noise with variance adjusted to obtain a target SNR level. If we let $\wt{\mu}_{m} \in \R$ and $\mu_{m} \in \R$ denote the $m$th true and assumed view angles, the view angle measurement coordinate uncertainty level is $\text{SNR}(\wt{\mu}_{m},\, \mu_{m})$.

Similarly, the assumed detector locations are uniformly spaced on the normalized interval $[0, 1]$ and the true detector locations can have an unknown perturbation from the assumed detector locations. To simulate detector uncertainty, we first independently perturb the first and last detectors from their assumed locations with a uniform perturbation scaled so that their location uncertainty meets a target SNR. The remaining detectors are then perturbed so that that the final true unknown detector array is uniformly spaced between the perturbed first and last detectors. Unlike the view angle measurement coordinate uncertainty model, this detector location coordinate uncertainty model is not iid. This enables us to evaluate the performance of our framework when there are non-iid measurement coordinate uncertainties.}

\rev{Combining these spaces gives} the measurement coordinates space as $\Omega = [0, \pi] \times [0, 1]$. We explore the performance of our method compared to the baseline which does not account for measurement coordinate uncertainty. We use images from the the LoDoPaB-CT tomography dataset resized to $128 \times 128$~\cite{leuschner2019lodopab}. From this dataset, 35,000 samples were used to train the image reconstruction 2D Unet $G_{\vmu}(\cdot)$. Test images from this dataset are used in this section to verify our framework.

\rev{
\subsection{View angle uncertainty}
\label{sec:view_angles}
}

\rev{
In the first group of experiments we consider the case where there is only view angle uncertainty and no uncertainty in the detector locations. As there is only uncertainty in the view angle dimension and not in the detector location dimension, we only optimize the view angle dimension of the measurement coordinates in~\eqref{eq:joint_objective}.}

\subsubsection{Combinations of measurement noise and view angle error} \label{sec:exp_xray_combos}

To determine how our framework performs under different settings, we consider different combinations of measurement noise and \rev{view angle measurement coordinate uncertainty}. We do trials over 25 different test images, and in each trial, different measurement noise and view angle perturbations are used.

\begin{figure}[t]
    \centering
    \subfloat[90 view angles
    \label{fig:image_sweep_90}]
    {\includegraphics[width=1\linewidth]{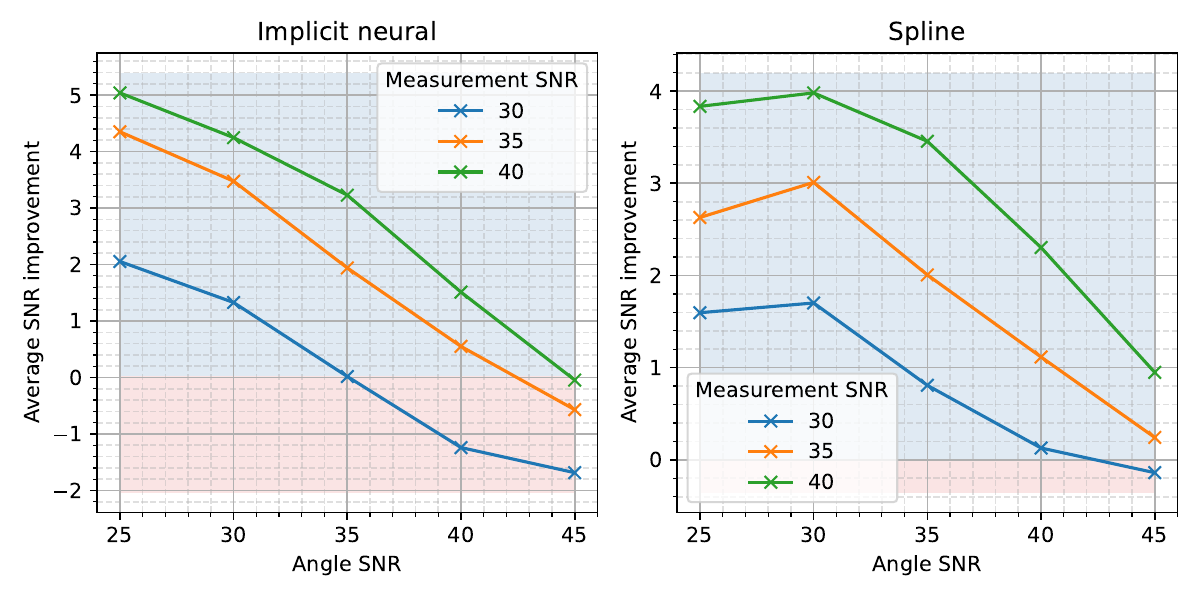}}
    
    \subfloat[120 view angles
    \label{fig:image_sweep_120}]
    {\includegraphics[width=1\linewidth]{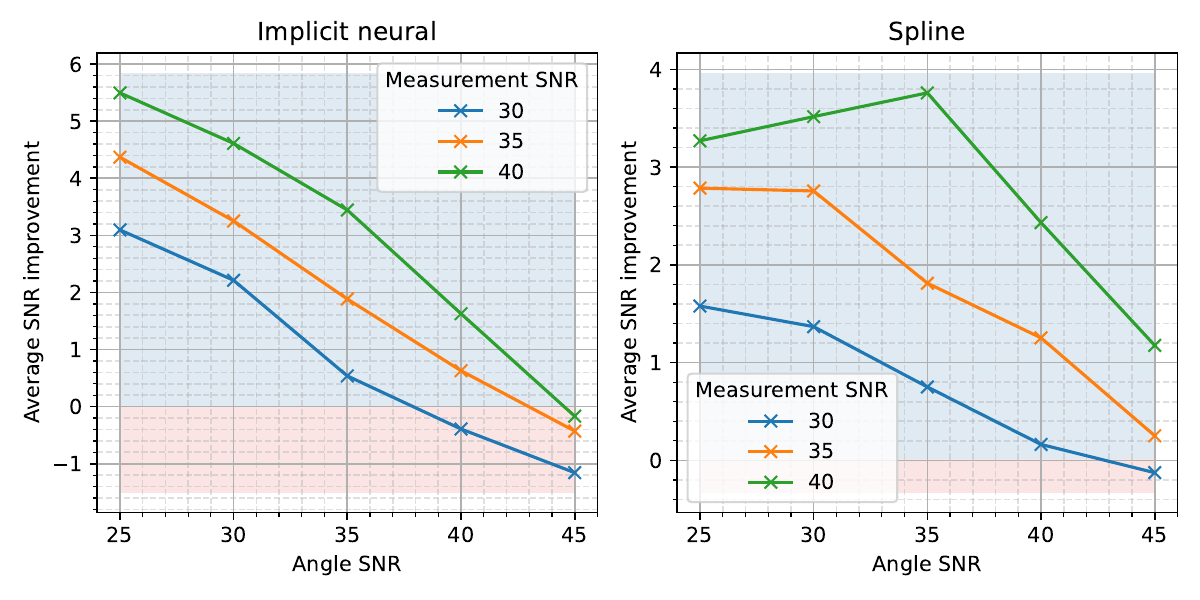}}
    
    \caption{SNR improvement (dB) when solving  \eqref{eq:joint_objective} \rev{for 2D CT imaging. There are }different combinations of measurement noise SNR and view angle uncertainty SNR. \rev{We consider} 90 and 120 view angles.}
    \label{fig:image_sweep}
\end{figure}

We evaluate the performance of our framework relative to the baseline by calculating the average reconstruction SNR improvement over the baseline. Fig. \ref{fig:image_sweep_90} shows the performance for 90 view angles. The performance trends are similar for both implicit neural and spline measurement representations. For a given measurement noise SNR, our method shows increasing improvements as the view angle SNR decreases (measurement coordinate uncertainty increases). This shows that our method handles measurement coordinate uncertainty well, especially when the uncertainties begin to dominate over measurement noise. 
We can also see that for a fixed angle SNR, the performance gains decrease as measurement SNR decreases and measurement errors becomes more dominant. Fig. \ref{fig:image_sweep_120} shows that the same trends hold when there are 120 view angles.
The slight SNR drops seen with splines in Fig. \ref{fig:image_sweep_90}, when the measurement SNR is 30 dB and 35 dB, do not appear in Fig. \ref{fig:image_sweep_120}. This is because there are more view angles which makes the imaging problem less ill-posed.

\begin{figure*}[t]
    \centering
    \includegraphics[width=0.9\linewidth]{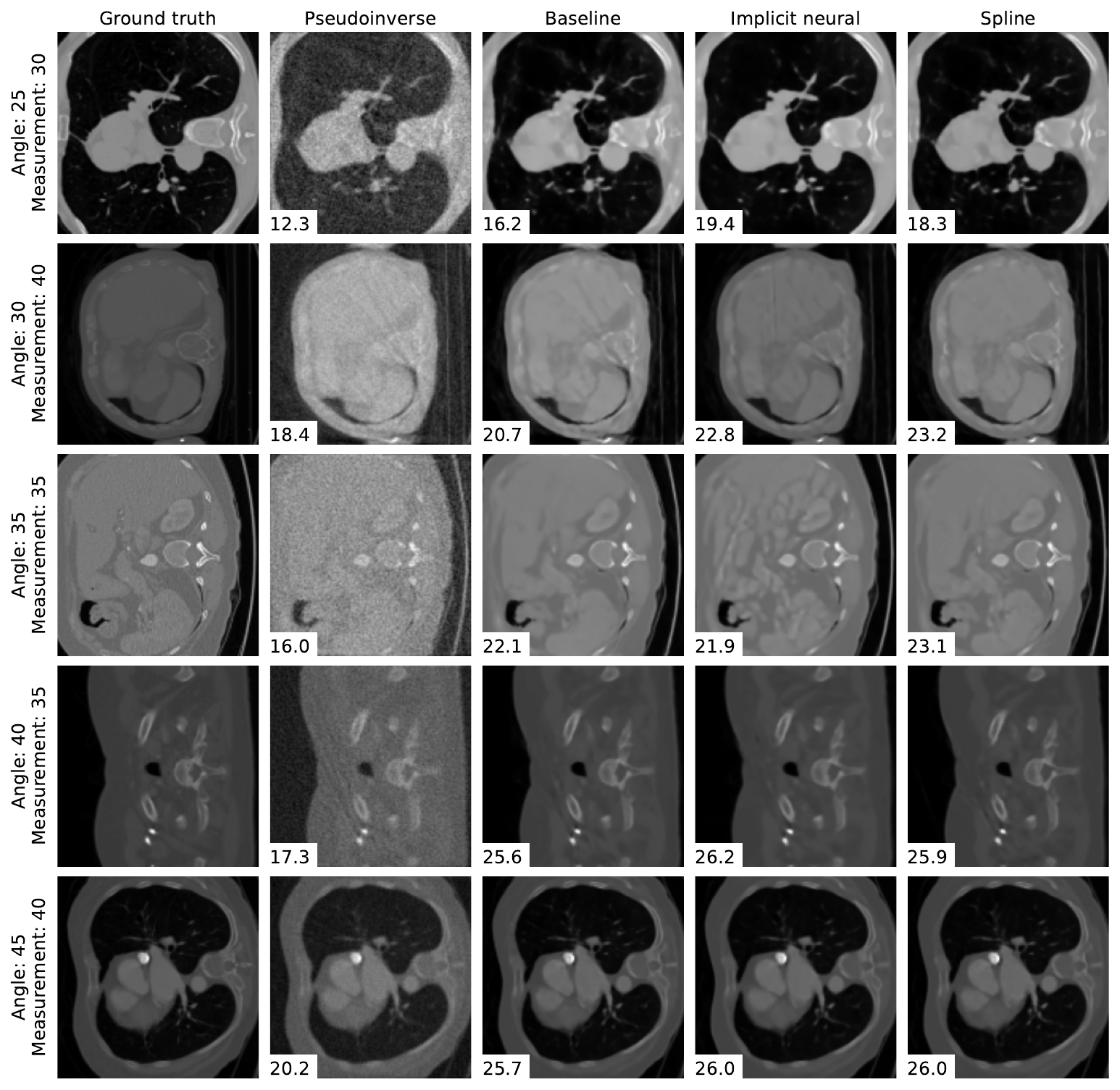}
    \caption{Example reconstructions for different measurement noise SNR and view angle uncertainty SNR combinations \rev{for 2D CT imaging. There} are 90 view angles. The reconstruction SNRs are shown for each reconstruction.}
    \label{fig:image_sweep_recons}
\end{figure*}

In Fig. \ref{fig:image_sweep_recons}, we show some \rev{randomly chosen} example ground truth, pseudoinverse filtered backprojection (FBP) and baseline reconstructions with their SNRs when there are 90 view angles. The reconstructions using our framework with implicit neural and spline measurement representations are also shown. Compared to the baseline, solutions obtained using our framework have fewer artifacts.
\rev{Note that Fig. \ref{fig:image_sweep_recons} shows specific examples, and that the average performance for different combinations of measurement noise and view angle uncertainty is shown in Fig. \ref{fig:image_sweep_90}.}
Fig. \ref{fig:image_sweep_recons_120} in the Appendix shows these same reconstructions when there are 120 view angles instead.

\subsubsection{Learned view angles accuracy} \label{sec:exp_xray_angle_error}

In the next \rev{numerical} experiment we verify that the learned measurement coordinates, $\wh{\vmu}$ in \eqref{eq:joint_objective}, are close to the true unknown measurement coordinates $\wt{\vmu}$. As the detector locations have no error, we verify the learned view angles only. We denote the set of true unknown view angles and learned view angles as $\wt{\Theta}$ and $\wh{\Theta}$. We quantitatively measure the average angle error \rev{in degrees} as 
\rev{
\begin{align}
    \text{Average angle error}
    =
    \frac{1}{J}
    \|{\wt{\Theta} - \wh{\Theta}}\|_1, \label{eq:angle_error}
\end{align}
where $J$ is the number of view angles.
}

\begin{figure}[t]
    \centering
    \includegraphics[width=0.9\linewidth]{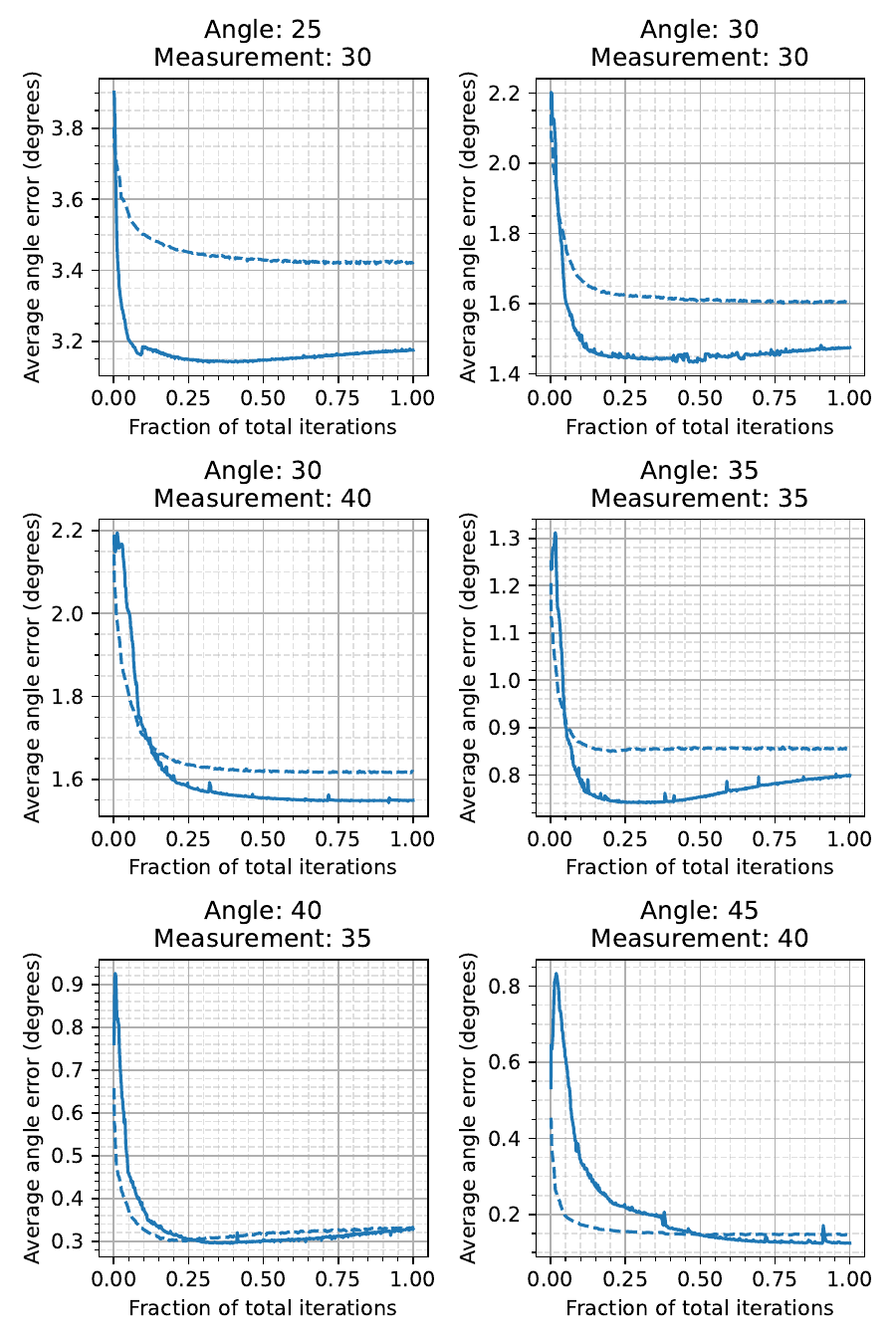}
    \caption{Average angle error for one test image with different combinations of measurement noise and view angle uncertainty when there are 90 \rev{2D CT} view angles. The solid lines are for implicit neural representations and the dashed lines are for spline representations.}
    \label{fig:image_sweep_angle_error}
\end{figure}

Fig. \ref{fig:image_sweep_angle_error} shows how the average angle error changes as the optimization iterations of \eqref{eq:joint_objective} progress. This is shown for one \rev{of the test images} with different combinations of measurement noise and \rev{view angle} uncertainty when there are 90 view angles. The solid lines are for implicit neural representations and the dashed lines are for spline representations. For both representation types, the angle error reduces as \eqref{eq:joint_objective} is solved which confirms that our framework learns measurement coordinates that are more accurate than the assumed measurement coordinates which they were initialized with. The average angle error when using the assumed measurement coordinates for reconstruction, as is done in the baseline, is the initial point on the plots.
\rev{In some instances, the average angle error may increase slightly as the optimization of \eqref{eq:joint_objective} progresses. This behavior can be explained as overfitting of the measurement representations.}
Fig. \ref{fig:image_sweep_angle_error_120} in the Appendix shows that the same trends hold when when there are 120 view angles.

\subsubsection{Reconstruction with more measurements}

Next we investigate a variant of the main problem considered in this paper: in addition to the $M$ true measurement coordinates being unknown, the reconstruction method $G_{\vmu}(\cdot)$ is now designed for $M'$ measurements where $M' \geq M$. In this case $\wt{\vmu} = (\wt{\mu}_1, \ldots, \wt{\mu}_M)$ as before and now $\vmu = (\mu_1, \ldots, \mu_{M'})$. As the measurement representation $r_{\vvarphi}(\cdot)$ can be evaluated at any measurement coordinate, we can evaluate \eqref{eq:loss_consistency} at $M'$ measurement coordinates.

In this \rev{simulated} experiment we obtain measurements from 90 view angles. As in the previous experiments, the true view angles are unknown and we initialize the angles of measurements coordinates $\vnu$ in \eqref{eq:loss_fitting} to be 90 uniformly spaced view angles in the interval $[0, \pi]$. The detector locations have no uncertainty. We consider different values of $M'$ by varying the number of view angles $J'$. The number of detectors $K$ are not varied which gives $M' = J' \cdot K$. \rev{Again, to obtain the strongest performance, we use state-of-the-art reconstruction Unets}. We try two different reconstruction Unets: 1) $G_{\vmu}^{J'}(\cdot)$ which was trained with training data having $J'$ view angles uniformly spaced on $[0, \pi]$ and, 2) $G_{\vmu}^{90}(\cdot)$ which was trained with training data having 90 view angles uniformly spaced on $[0, \pi]$.\footnote{The number of Unet training view angles and evaluated view angles $J'$ can be different. This is because the Unet input is computed by a filtered backprojection for $J'$ view angles which always results in an Unet input with the dimensions of the image being reconstructed.} 

Table \ref{table:interpolation} shows the average reconstruction SNR over 25 test images. There is 35 dB measurement noise and the unknown true view angles are perturbed by 35 dB from 90 uniformly spaced view angles. With Unet $G_{\mu}^{J'}(\cdot)$, the performance fluctuates. The performance is stable with Unet $G_{\vmu}^{90}(\cdot)$. 
With both reconstruction methods, the best performance is seen when $J' = 90$ ($M' = M$) and evaluating more measurements does not help. This is because the fitting loss \eqref{eq:loss_fitting} ensures that $r_{\vvarphi}(\cdot)$ accurately represents the $M$ obtained measurements which were sampled from the measurement space at coordinates $\wt{\vmu}$.
Then when $M' = M$, because the measurement coordinates used for reconstruction, $\vmu$, are a small perturbation away from $\wt{\vmu}$, they are also represented accurately which results in good reconstructions.
Furthermore, when $M' = M$, there are enough measurement coordinates that densely cover the measurement space. The accuracy of the measurement coordinates for the extra measurements when $M' > M$ is not enforced by the fitting loss \eqref{eq:loss_fitting}.

\begin{table}[t]
\caption{Average \rev{2D CT} reconstruction SNR when reconstructing with more measurements.}
\label{table:interpolation}
\centering
\begin{tabular}{@{}ccccc@{}}
\toprule
& \multicolumn{2}{c}{$G_{\vmu}^{J'}(\cdot)$} & \multicolumn{2}{c}{$G_{\vmu}^{90}(\cdot)$} \\
$J'$ & Implicit & Spline & Implicit & Spline \\
\midrule
90  & \textbf{23.5}    & \textbf{23.6}    & \textbf{23.5}    & \textbf{23.6} \\
110 & 22.6    & 23.1    & \textbf{23.7}    & \textbf{23.6} \\
120 & 23.4    & 23.1    & \textbf{23.8}    & \textbf{23.6} \\
130 & 23.2    & 22.8    & \textbf{23.7}    & \textbf{23.5} \\
150 & 22.9    & 22.5    & \textbf{23.8}    & \textbf{23.6} \\
180 & 22.3    & 22.0    & \textbf{23.9}    & \textbf{23.6} \\
\bottomrule
\end{tabular}
\end{table}

It has been shown that reconstruction with more measurements can help when there is no measurement coordinate uncertainty, and when the input to the reconstruction neural network combines the observed measurements with the measurement representation output \cite{sun2021coil}. When there is measurement coordinate uncertainty, using the observed measurements in the input to the reconstruction neural network can reduce performance as shown by Fig. \ref{fig:parameter_mismatch} and the baseline reconstructions in Fig. \ref{fig:image_sweep_recons}.

\rev{
\subsection{Detector location uncertainty}
\label{sec:detector_uncertainty}

The previous experiments demonstrate that our framework improves over the baseline when there is uncertainty in the view angles. Next we consider the case where there is only uncertainty in the detector location and no uncertainty in the view angles. In this set of experiments there are always 90 uniformly spaced view angles. As there is only uncertainty in detector locations, we only optimize the detector location dimension of the measurement coordinates in~\eqref{eq:joint_objective}.

\subsubsection{Combinations of measurement noise and detector location error}

Similar to Section \ref{sec:exp_xray_combos}, we consider different combinations of measurement noise and detector location uncertainty. We do trials over the same 25 test images. In each trial, different measurement noise and detector location perturbations are used.

The average reconstruction SNR improvement over the baseline is shown in Fig. \ref{fig:image_sweep_two_coords_no_angle_error}. For a given measurement noise SNR, our method shows increasing improvements as the detector location SNR decreases (measurement coordinate uncertainty increases). This is consistent with the previous results for view angle measurement coordinate uncertainty shown in Fig.~\ref{fig:image_sweep}. Our method handles measurement coordinate uncertainty well, especially when the uncertainties begin to dominate over measurement noise. \aqe{Note that our framework does not assume a specific uncertainty model---we have used different uncertainty simulation models for view angle uncertainty in Fig. \ref{fig:image_sweep} and detector location uncertainty in Fig. \ref{fig:image_sweep_two_coords_no_angle_error}. Compared to Fig. \ref{fig:image_sweep}, the curves in Fig. \ref{fig:image_sweep_two_coords_no_angle_error} for high measurement SNR and measurement coordinate uncertainty SNR almost overlap at times because the baseline SNR is higher and so the average SNR improvement is lower.} 

\begin{figure}[t]
    \centering
    \includegraphics[width=1\linewidth]{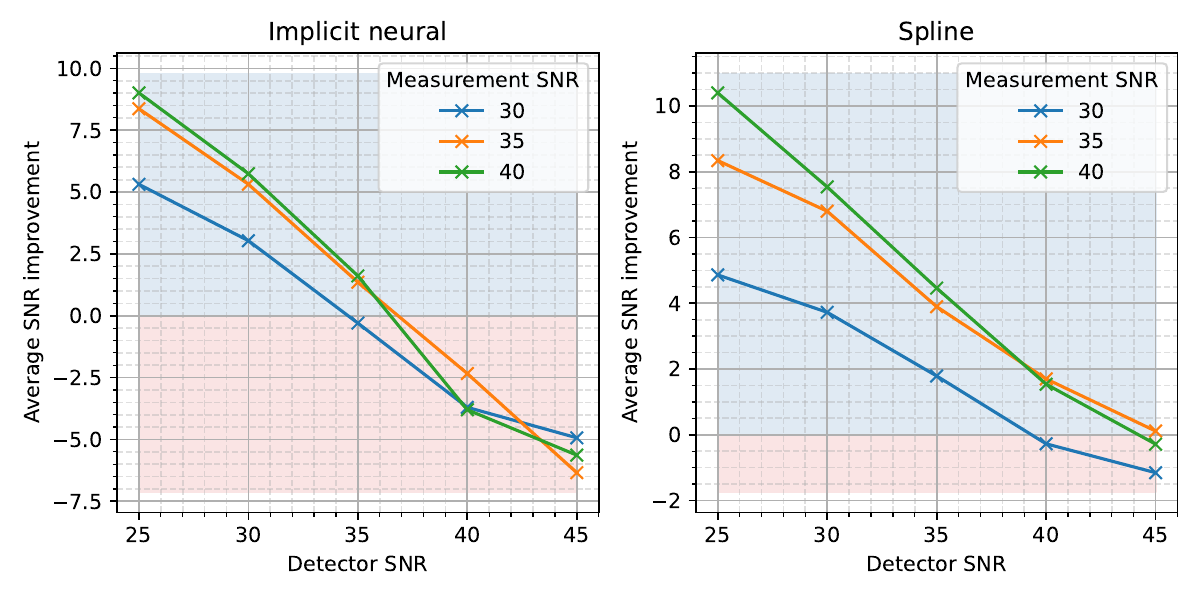}
    \caption{Average SNR improvement (dB) when solving \eqref{eq:joint_objective} for 2D CT imaging. There are different combinations of measurement noise SNR and detector location uncertainty SNR. There are 90 view angles.}
    \label{fig:image_sweep_two_coords_no_angle_error}
\end{figure}

In Fig. \ref{fig:image_sweep_recons_detector}, we show randomly chosen example reconstructions with their SNRs. 

\begin{figure*}[t]
    \centering
    \includegraphics[width=0.9\linewidth]{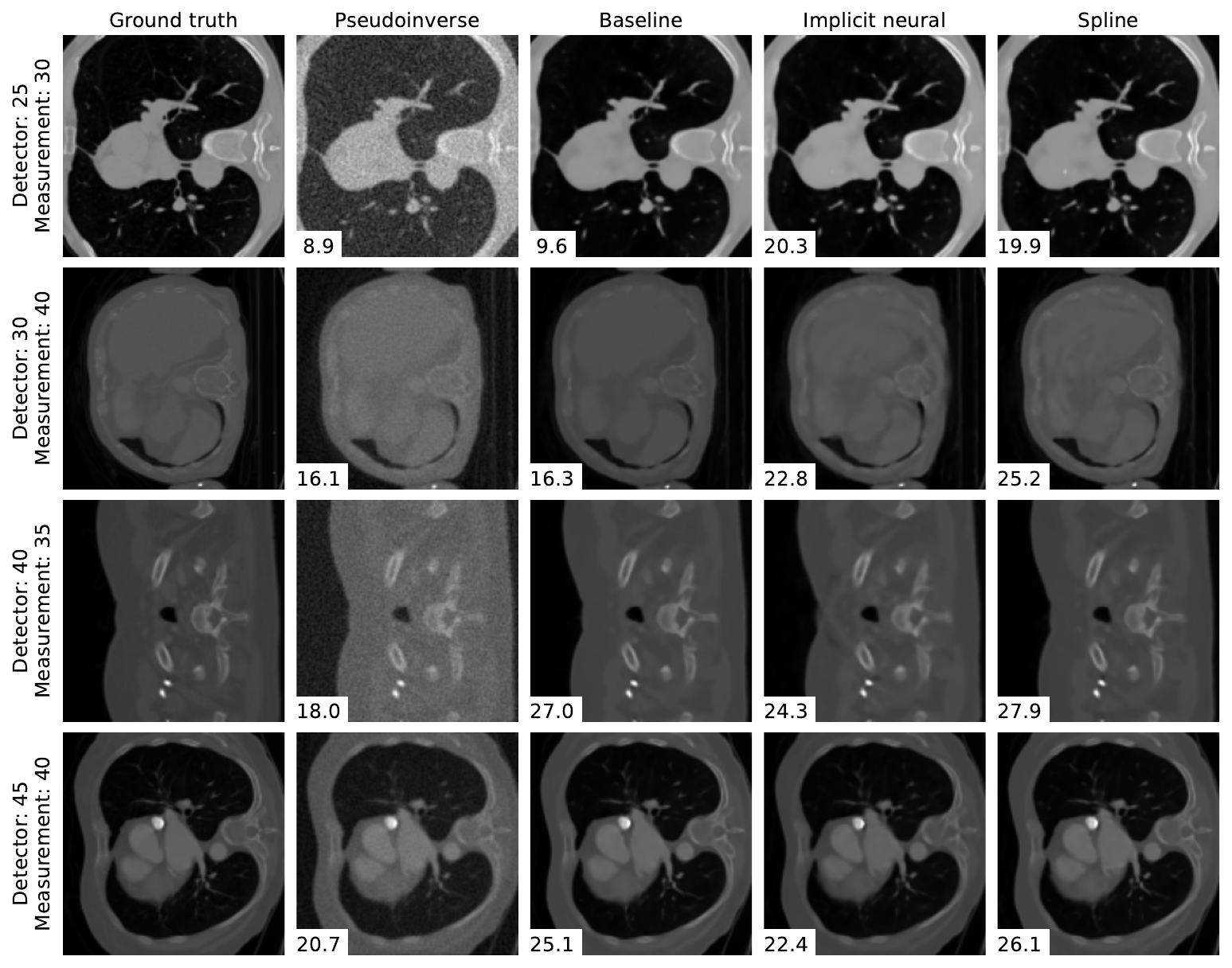}
    \caption{Example reconstructions for 2D CT imaging. There are different measurement noise SNR and detector location uncertainty SNR combinations and 90 view angles. The reconstruction SNRs are shown for each reconstruction.}
    \label{fig:image_sweep_recons_detector}
\end{figure*}

\subsubsection{Learned detector locations accuracy}

Similar to Section \ref{sec:exp_xray_angle_error}, in the next simulation we verify that the learned detector locations are close to the true unknown detector locations. We denote the set of true unknown detector locations and learned detector locations as $\wt{\calT}$ and $\wh{\calT}$. We measure the average detector location error as
\begin{align}
    \text{Average detector error}
    =
    \frac{1}{K}
    \|{\wt{\calT} - \wh{\calT}}\|_1,
\end{align}
where $K$ is the number of detectors.

Fig. \ref{fig:image_sweep_detector_error} shows how the average detector error changes as the optimization progresses for different combinations of measurement noise and detector location uncertainty. Again, the solid lines are for implicit neural representations and the dashed lines are for spline representations. Compared to the assumed detector locations, for both representation types, the final learned detector locations are closer to the true detector locations.

\begin{figure}[t]
    \centering
    \includegraphics[width=0.9\linewidth]{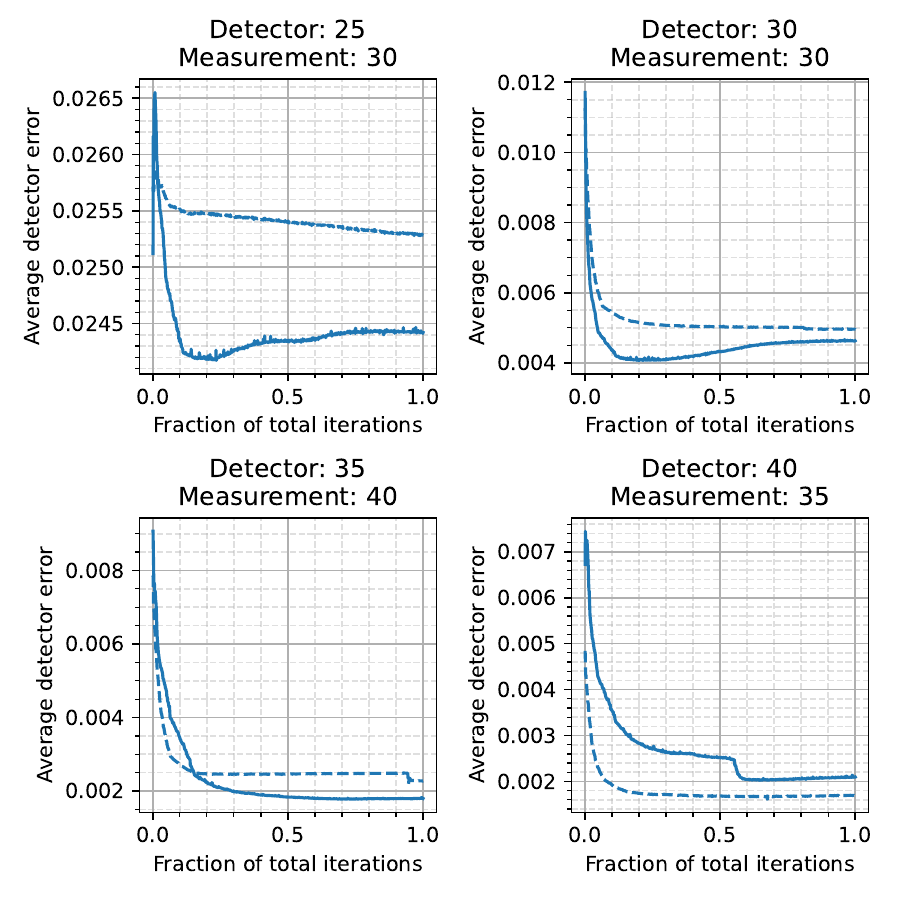}
    \caption{Average detector error for one test image with different combinations of measurement noise and measurement detector location uncertainty when performing 2D CT imaging. The solid lines are for implicit neural representations and the dashed lines are for spline representations.}
    \label{fig:image_sweep_detector_error}
\end{figure}

\subsection{View angle and detector location uncertainty}

The numerical experiments in Sections \ref{sec:view_angles} and \ref{sec:detector_uncertainty} demonstrates that our proposed framework performs well when there is view angle or detector location measurement coordinate uncertainty. We now perform simulations when there is view angle and detector location uncertainty at the same time. There are 90 view angles in these experiments. As there is uncertainty in both the view angles and the detector locations, we will optimize both the view angle and detector location dimensions of the measurement coordinates in~\eqref{eq:joint_objective}.

\subsubsection{Combinations of measurement noise and measurement coordinate error}

Following the experiments of Figs. \ref{fig:image_sweep} and \ref{fig:image_sweep_two_coords_no_angle_error}, we consider different combinations of measurement noise and measurement coordinate uncertainty over the same 25 test images. The view angle and detector location SNRs are the same in these experiments.

The average reconstruction SNR improvement over the baseline is shown in Fig. \ref{fig:image_sweep_two_coords_same_error}. \aqe{Consistent with previous experiments, Fig. \ref{fig:image_sweep_two_coords_same_error} shows that we are able to improve upon the baseline when there is measurement coordinate uncertainty.}
In Fig. \ref{fig:image_sweep_recons_two_coords}, we also show randomly chosen example reconstructions with their SNRs when there is measurement coordinate uncertainty in both the view angles and the detector locations.

\begin{figure}[t]
    \centering
    \includegraphics[width=1\linewidth]{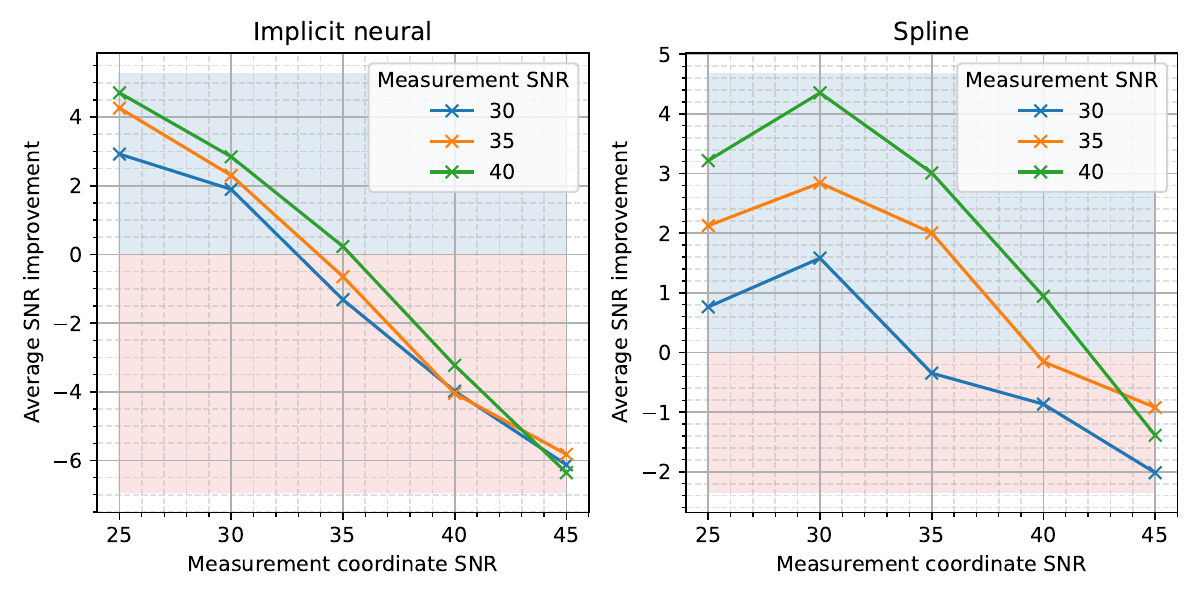}
    \caption{Average SNR improvement (dB) for 2D CT imaging when solving \eqref{eq:joint_objective} for different combinations of measurement noise SNR and measurement coordinate uncertainty. There is uncertainty in both the view angles and detector locations.}
    \label{fig:image_sweep_two_coords_same_error}
\end{figure}

\begin{figure*}[t]
    \centering
    \includegraphics[width=0.9\linewidth]{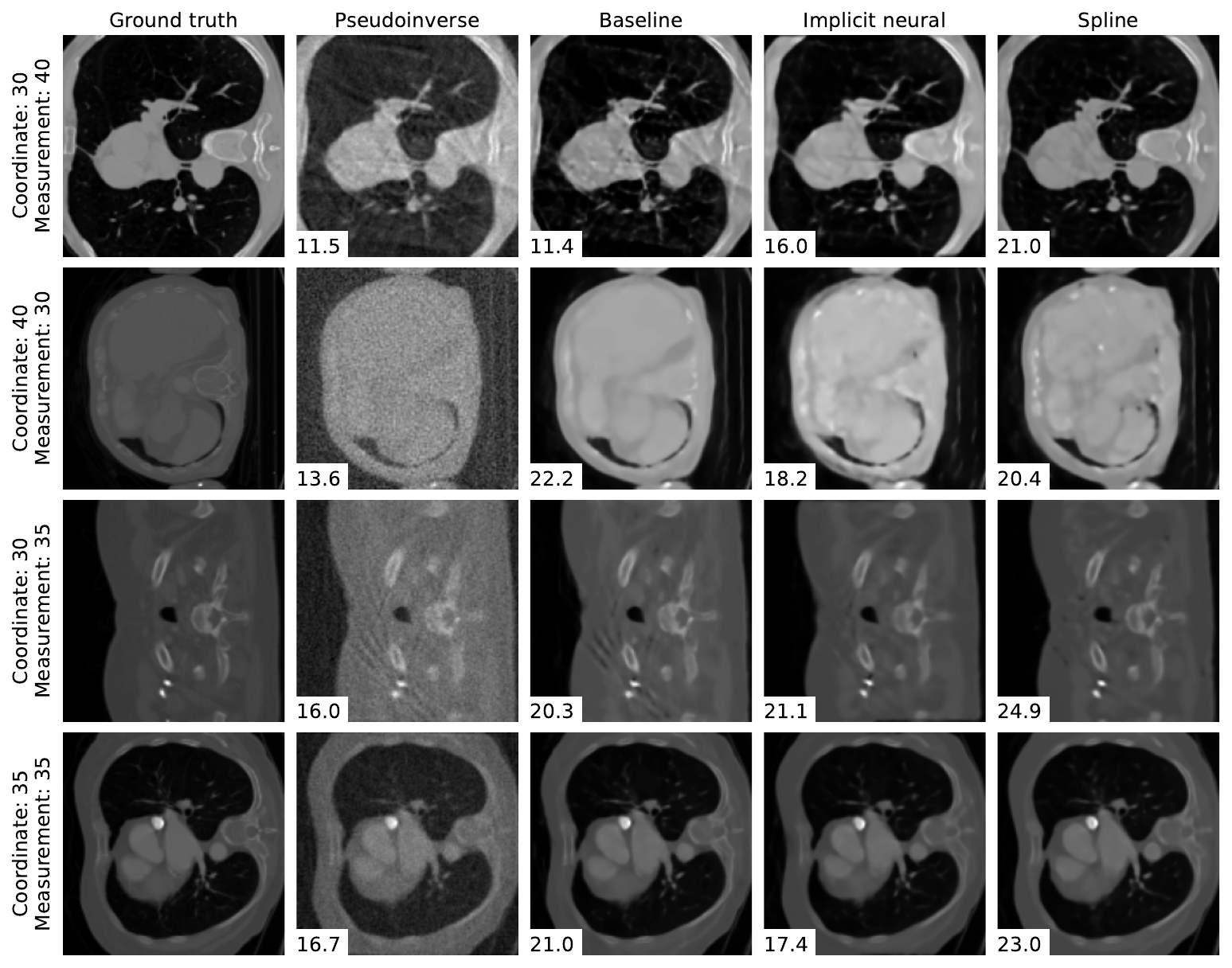}
    \caption{Example reconstructions for 2D CT imaging. There are different measurement noise SNR and measurement coordinate uncertainty SNR combinations. There is uncertainty in both the view angles and detector locations. The reconstruction SNRs are shown for each reconstruction.}
    \label{fig:image_sweep_recons_two_coords}
\end{figure*}

}

\section{Conclusion} \label{sec:conclusion}

We presented a differentiable imaging inverse problem framework to jointly reconstruct the unknown image and learn unknown measurement coordinates when they are approximately known. There are two major elements in our proposed method. Firstly, we learn continuous representations of the measurements whose input are measurement coordinates and output are the corresponding measurements. By optimizing with respect to their parameters and their input, we jointly learn the measurement representation parameters and the unknown measurement coordinates. The second aspect of our method is that because these representations can be evaluated at any input coordinate, we can leverage reconstruction methods that are designed for measurement coordinates that are different from the ones of the observations.
Our 2D and 3D CT imaging \rev{simulations} show that \rev{the benefit of using our framework increases with the level of measurement coordinate uncertainty.}

As our framework does not assume a particular measurement representation, we use both implicit neural networks and splines to represent measurements. Splines are generally viewed as interpolation tools, however, our work demonstrates that they can also be learnable differentiable representations that perform comparably to implicit neural representations. Differentiable splines may provide a viable solution for current research directions that have been focusing on using implicit neural networks which can have significantly more parameters and complexity \cite{xie2022neural}.

A strength of our framework is that no extra \rev{training} data is required to learn the measurement representations. However, a drawback is that it can be time consuming if there are multiple test images because we have to learn separate measurement representations and measurement coordinates for each new set of observations. Therefore, extending our framework to jointly recover a batch of images and the shared unknown measurement coordinates is an important step towards helping practitioners adopt our framework. Batch imaging also introduces robustness which can help learn more accurate measurement representations.

Another important endeavor is to adapt our framework to account for operator uncertainties due to reasons other than measurement coordinate uncertainty which this paper studied. For example, there may be approximation uncertainties if the true operator is approximated to enable faster computations and facilitate analysis. Not accounting for the approximation can lead to degraded solutions \cite{lunz2021learned}. Another source of operator uncertainty arises when the object being imaged is altered in an unknown manner during the measurement acquisition process. For example, in CryoET imaging, the sample can translate and deform during imaging which needs to be taken into account~\cite{tegunov2019real, naydenova2020cryo, fernandez2021tomoalign}.

\bibliographystyle{IEEEtran}
\bibliography{refs}

\begin{thebibliography}{10}
\providecommand{\url}[1]{#1}
\csname url@samestyle\endcsname
\providecommand{\newblock}{\relax}
\providecommand{\bibinfo}[2]{#2}
\providecommand{\BIBentrySTDinterwordspacing}{\spaceskip=0pt\relax}
\providecommand{\BIBentryALTinterwordstretchfactor}{4}
\providecommand{\BIBentryALTinterwordspacing}{\spaceskip=\fontdimen2\font plus
\BIBentryALTinterwordstretchfactor\fontdimen3\font minus
  \fontdimen4\font\relax}
\providecommand{\BIBforeignlanguage}[2]{{%
\expandafter\ifx\csname l@#1\endcsname\relax
\typeout{** WARNING: IEEEtran.bst: No hyphenation pattern has been}%
\typeout{** loaded for the language `#1'. Using the pattern for}%
\typeout{** the default language instead.}%
\else
\language=\csname l@#1\endcsname
\fi
#2}}
\providecommand{\BIBdecl}{\relax}
\BIBdecl

\bibitem{ronneberger2015u}
O.~Ronneberger, P.~Fischer, and T.~Brox, ``U-net: Convolutional networks for
  biomedical image segmentation,'' in \emph{International Conference on Medical
  image computing and computer-assisted intervention}.\hskip 1em plus 0.5em
  minus 0.4em\relax Springer, 2015, pp. 234--241.

\bibitem{jin2017deep}
K.~H. Jin, M.~T. McCann, E.~Froustey, and M.~Unser, ``Deep convolutional neural
  network for inverse problems in imaging,'' \emph{IEEE Transactions on Image
  Processing}, vol.~26, no.~9, pp. 4509--4522, 2017.

\bibitem{mccann2017convolutional}
M.~T. McCann, K.~H. Jin, and M.~Unser, ``Convolutional neural networks for
  inverse problems in imaging: A review,'' \emph{IEEE Signal Processing
  Magazine}, vol.~34, no.~6, pp. 85--95, 2017.

\bibitem{ongie2020deep}
G.~Ongie, A.~Jalal, C.~A. Metzler, R.~G. Baraniuk, A.~G. Dimakis, and
  R.~Willett, ``Deep learning techniques for inverse problems in imaging,''
  \emph{IEEE Journal on Selected Areas in Information Theory}, vol.~1, no.~1,
  pp. 39--56, 2020.

\bibitem{antholzer2019deep}
S.~Antholzer, M.~Haltmeier, and J.~Schwab, ``Deep learning for photoacoustic
  tomography from sparse data,'' \emph{Inverse problems in science and
  engineering}, vol.~27, no.~7, pp. 987--1005, 2019.

\bibitem{kothari2018random}
K.~Kothari, S.~Gupta, M.~v.~de Hoop, and I.~Dokmanic, ``Random mesh projectors
  for inverse problems,'' in \emph{International Conference on Learning
  Representations}, 2019.

\bibitem{ulyanov2018deep}
D.~Ulyanov, A.~Vedaldi, and V.~Lempitsky, ``Deep image prior,'' in
  \emph{Proceedings of the IEEE conference on computer vision and pattern
  recognition}, 2018, pp. 9446--9454.

\bibitem{gregor2010learning}
K.~Gregor and Y.~LeCun, ``Learning fast approximations of sparse coding,'' in
  \emph{Proceedings of the 27th international conference on international
  conference on machine learning}, 2010, pp. 399--406.

\bibitem{adler2018learned}
J.~Adler and O.~{\"O}ktem, ``Learned primal-dual reconstruction,'' \emph{IEEE
  transactions on medical imaging}, vol.~37, no.~6, pp. 1322--1332, 2018.

\bibitem{gupta2018cnn}
H.~Gupta, K.~H. Jin, H.~Q. Nguyen, M.~T. McCann, and M.~Unser, ``Cnn-based
  projected gradient descent for consistent ct image reconstruction,''
  \emph{IEEE transactions on medical imaging}, vol.~37, no.~6, pp. 1440--1453,
  2018.

\bibitem{rick2017one}
J.~Rick~Chang, C.-L. Li, B.~Poczos, B.~Vijaya~Kumar, and A.~C.
  Sankaranarayanan, ``One network to solve them all--solving linear inverse
  problems using deep projection models,'' in \emph{Proceedings of the IEEE
  International Conference on Computer Vision}, 2017, pp. 5888--5897.

\bibitem{gilton2019neumann}
D.~Gilton, G.~Ongie, and R.~Willett, ``Neumann networks for linear inverse
  problems in imaging,'' \emph{IEEE Transactions on Computational Imaging},
  vol.~6, pp. 328--343, 2019.

\bibitem{zhang2021plug}
K.~Zhang, Y.~Li, W.~Zuo, L.~Zhang, L.~Van~Gool, and R.~Timofte, ``Plug-and-play
  image restoration with deep denoiser prior,'' \emph{IEEE Transactions on
  Pattern Analysis and Machine Intelligence}, 2021.

\bibitem{bora2017compressed}
A.~Bora, A.~Jalal, E.~Price, and A.~G. Dimakis, ``Compressed sensing using
  generative models,'' in \emph{International Conference on Machine
  Learning}.\hskip 1em plus 0.5em minus 0.4em\relax PMLR, 2017, pp. 537--546.

\bibitem{kothari2021trumpets}
K.~Kothari, A.~Khorashadizadeh, M.~de~Hoop, and I.~Dokmani{\'c}, ``Trumpets:
  Injective flows for inference and inverse problems,'' in \emph{Uncertainty in
  Artificial Intelligence}.\hskip 1em plus 0.5em minus 0.4em\relax PMLR, 2021,
  pp. 1269--1278.

\bibitem{song2021solving}
Y.~Song, L.~Shen, L.~Xing, and S.~Ermon, ``Solving inverse problems in medical
  imaging with score-based generative models,'' \emph{arXiv preprint
  arXiv:2111.08005}, 2021.

\bibitem{gilton2021model}
D.~Gilton, G.~Ongie, and R.~Willett, ``Model adaptation for inverse problems in
  imaging,'' \emph{IEEE Transactions on Computational Imaging}, vol.~7, pp.
  661--674, 2021.

\bibitem{gossard2022training}
A.~Gossard and P.~Weiss, ``Training adaptive reconstruction networks for
  inverse problems,'' \emph{arXiv preprint arXiv:2202.11342}, 2022.

\bibitem{bartels1995introduction}
R.~H. Bartels, J.~C. Beatty, and B.~A. Barsky, \emph{An introduction to splines
  for use in computer graphics and geometric modeling}.\hskip 1em plus 0.5em
  minus 0.4em\relax Morgan Kaufmann, 1995.

\bibitem{rogers1989mathematical}
D.~F. Rogers and J.~A. Adams, \emph{Mathematical elements for computer
  graphics}.\hskip 1em plus 0.5em minus 0.4em\relax McGraw-Hill, Inc., 1989.

\bibitem{piegl1991nurbs}
L.~Piegl, ``On nurbs: a survey,'' \emph{IEEE Computer Graphics and
  Applications}, vol.~11, no.~01, pp. 55--71, 1991.

\bibitem{cohen1980discrete}
E.~Cohen, T.~Lyche, and R.~Riesenfeld, ``Discrete b-splines and subdivision
  techniques in computer-aided geometric design and computer graphics,''
  \emph{Computer graphics and image processing}, vol.~14, no.~2, pp. 87--111,
  1980.

\bibitem{prasad2022nurbs}
A.~D. Prasad, A.~Balu, H.~Shah, S.~Sarkar, C.~Hegde, and A.~Krishnamurthy,
  ``Nurbs-diff: A differentiable programming module for nurbs,''
  \emph{Computer-Aided Design}, vol. 146, p. 103199, 2022.

\bibitem{xie2022neural}
Y.~Xie, T.~Takikawa, S.~Saito, O.~Litany, S.~Yan, N.~Khan, F.~Tombari,
  J.~Tompkin, V.~Sitzmann, and S.~Sridhar, ``Neural fields in visual computing
  and beyond,'' in \emph{Computer Graphics Forum}, vol.~41, no.~2.\hskip 1em
  plus 0.5em minus 0.4em\relax Wiley Online Library, 2022, pp. 641--676.

\bibitem{park2019deepsdf}
J.~J. Park, P.~Florence, J.~Straub, R.~Newcombe, and S.~Lovegrove, ``Deepsdf:
  Learning continuous signed distance functions for shape representation,'' in
  \emph{Proceedings of the IEEE/CVF conference on computer vision and pattern
  recognition}, 2019, pp. 165--174.

\bibitem{mildenhall2020nerf}
B.~Mildenhall, P.~P. Srinivasan, M.~Tancik, J.~T. Barron, R.~Ramamoorthi, and
  R.~Ng, ``Nerf: Representing scenes as neural radiance fields for view
  synthesis,'' in \emph{European conference on computer vision}.\hskip 1em plus
  0.5em minus 0.4em\relax Springer, 2020, pp. 405--421.

\bibitem{sitzmann2020implicit}
V.~Sitzmann, J.~Martel, A.~Bergman, D.~Lindell, and G.~Wetzstein, ``Implicit
  neural representations with periodic activation functions,'' \emph{Advances
  in Neural Information Processing Systems}, vol.~33, pp. 7462--7473, 2020.

\bibitem{Reed_2021_ICCV}
A.~W. Reed, H.~Kim, R.~Anirudh, K.~A. Mohan, K.~Champley, J.~Kang, and
  S.~Jayasuriya, ``Dynamic ct reconstruction from limited views with implicit
  neural representations and parametric motion fields,'' in \emph{Proceedings
  of the IEEE/CVF International Conference on Computer Vision (ICCV)}, October
  2021, pp. 2258--2268.

\bibitem{lozenski2022neural}
L.~Lozenski, M.~Anastasio, and U.~Villa, ``Neural fields for dynamic imaging,''
  in \emph{Medical Imaging 2022: Physics of Medical Imaging}, vol. 12031.\hskip
  1em plus 0.5em minus 0.4em\relax SPIE, 2022, pp. 231--238.

\bibitem{sun2021coil}
Y.~Sun, J.~Liu, M.~Xie, B.~Wohlberg, and U.~S. Kamilov, ``Coil:
  Coordinate-based internal learning for tomographic imaging,'' \emph{IEEE
  Transactions on Computational Imaging}, vol.~7, pp. 1400--1412, 2021.

\bibitem{chen2023differentiable}
N.~Chen, L.~Cao, T.-C. Poon, B.~Lee, and E.~Y. Lam, ``Differentiable imaging: A
  new tool for computational optical imaging,'' \emph{Advanced Physics
  Research}, 2023.

\bibitem{du2020three}
M.~Du, Y.~S. Nashed, S.~Kandel, D.~G{\"u}rsoy, and C.~Jacobsen, ``Three
  dimensions, two microscopes, one code: Automatic differentiation for x-ray
  nanotomography beyond the depth of focus limit,'' \emph{Science advances},
  vol.~6, no.~13, p. eaay3700, 2020.

\bibitem{du2021adorym}
M.~Du, S.~Kandel, J.~Deng, X.~Huang, A.~Demortiere, T.~T. Nguyen, R.~Tucoulou,
  V.~De~Andrade, Q.~Jin, and C.~Jacobsen, ``Adorym: A multi-platform generic
  x-ray image reconstruction framework based on automatic differentiation,''
  \emph{Optics express}, vol.~29, no.~7, pp. 10\,000--10\,035, 2021.

\bibitem{triggs2000bundle}
B.~Triggs, P.~F. McLauchlan, R.~I. Hartley, and A.~W. Fitzgibbon, ``Bundle
  adjustment—a modern synthesis,'' in \emph{Vision Algorithms: Theory and
  Practice: International Workshop on Vision Algorithms Corfu, Greece,
  September 21--22, 1999 Proceedings}.\hskip 1em plus 0.5em minus 0.4em\relax
  Springer, 2000, pp. 298--372.

\bibitem{campisi2017blind}
P.~Campisi and K.~Egiazarian, \emph{Blind image deconvolution: theory and
  applications}.\hskip 1em plus 0.5em minus 0.4em\relax CRC press, 2017.

\bibitem{chan1998total}
T.~F. Chan and C.-K. Wong, ``Total variation blind deconvolution,'' \emph{IEEE
  transactions on Image Processing}, vol.~7, no.~3, pp. 370--375, 1998.

\bibitem{riis2021computed}
N.~A.~B. Riis, Y.~Dong, and P.~C. Hansen, ``Computed tomography reconstruction
  with uncertain view angles by iteratively updated model discrepancy,''
  \emph{Journal of Mathematical Imaging and Vision}, vol.~63, no.~2, pp.
  133--143, 2021.

\bibitem{basu2000uniqueness}
S.~Basu and Y.~Bresler, ``Uniqueness of tomography with unknown view angles,''
  \emph{IEEE Transactions on Image Processing}, vol.~9, no.~6, pp. 1094--1106,
  2000.

\bibitem{coifman2008graph}
R.~R. Coifman, Y.~Shkolnisky, F.~J. Sigworth, and A.~Singer, ``Graph laplacian
  tomography from unknown random projections,'' \emph{IEEE Transactions on
  Image Processing}, vol.~17, no.~10, pp. 1891--1899, 2008.

\bibitem{bendory2020single}
T.~Bendory, A.~Bartesaghi, and A.~Singer, ``Single-particle cryo-electron
  microscopy: Mathematical theory, computational challenges, and
  opportunities,'' \emph{IEEE signal processing magazine}, vol.~37, no.~2, pp.
  58--76, 2020.

\bibitem{golub1980analysis}
G.~H. Golub and C.~F. Van~Loan, ``An analysis of the total least squares
  problem,'' \emph{SIAM journal on numerical analysis}, vol.~17, no.~6, pp.
  883--893, 1980.

\bibitem{markovsky2007overview}
I.~Markovsky and S.~Van~Huffel, ``Overview of total least-squares methods,''
  \emph{Signal processing}, vol.~87, no.~10, pp. 2283--2302, 2007.

\bibitem{gupta2021total}
S.~Gupta and I.~Dokmani{\'c}, ``Total least squares phase retrieval,''
  \emph{IEEE Transactions on Signal Processing}, vol.~70, pp. 536--549, 2021.

\bibitem{piegl1996nurbs}
L.~Piegl and W.~Tiller, \emph{The NURBS book}.\hskip 1em plus 0.5em minus
  0.4em\relax Springer Science \& Business Media, 1996.

\bibitem{rogers2001introduction}
D.~F. Rogers, \emph{An introduction to NURBS: with historical
  perspective}.\hskip 1em plus 0.5em minus 0.4em\relax Morgan Kaufmann, 2001.

\bibitem{adler2017operator}
J.~Adler, H.~Kohr, and O.~{\"O}ktem, ``Operator discretization library (odl),''
  \emph{Zenodo}, 2017.

\bibitem{cciccek20163d}
{\"O}.~{\c{C}}i{\c{c}}ek, A.~Abdulkadir, S.~S. Lienkamp, T.~Brox, and
  O.~Ronneberger, ``3d u-net: learning dense volumetric segmentation from
  sparse annotation,'' in \emph{International conference on medical image
  computing and computer-assisted intervention}.\hskip 1em plus 0.5em minus
  0.4em\relax Springer, 2016, pp. 424--432.

\bibitem{leuschner2019lodopab}
J.~Leuschner, M.~Schmidt, D.~O. Baguer, and P.~Maa{\ss}, ``The lodopab-ct
  dataset: A benchmark dataset for low-dose ct reconstruction methods,''
  \emph{arXiv preprint arXiv:1910.01113}, 2019.

\bibitem{lunz2021learned}
S.~Lunz, A.~Hauptmann, T.~Tarvainen, C.-B. Schonlieb, and S.~Arridge, ``On
  learned operator correction in inverse problems,'' \emph{SIAM Journal on
  Imaging Sciences}, vol.~14, no.~1, pp. 92--127, 2021.

\bibitem{tegunov2019real}
D.~Tegunov and P.~Cramer, ``Real-time cryo-electron microscopy data
  preprocessing with warp,'' \emph{Nature methods}, vol.~16, no.~11, pp.
  1146--1152, 2019.

\bibitem{naydenova2020cryo}
K.~Naydenova, P.~Jia, and C.~J. Russo, ``Cryo-em with sub--1 {\aa} specimen
  movement,'' \emph{Science}, vol. 370, no. 6513, pp. 223--226, 2020.

\bibitem{fernandez2021tomoalign}
J.-J. Fernandez and S.~Li, ``Tomoalign: A novel approach to correcting sample
  motion and 3d ctf in cryoet,'' \emph{Journal of Structural Biology}, vol.
  213, no.~4, p. 107778, 2021.

\bibitem{menze2014multimodal}
B.~H. Menze, A.~Jakab, S.~Bauer, J.~Kalpathy-Cramer, K.~Farahani, J.~Kirby,
  Y.~Burren, N.~Porz, J.~Slotboom, R.~Wiest \emph{et~al.}, ``The multimodal
  brain tumor image segmentation benchmark (brats),'' \emph{IEEE transactions
  on medical imaging}, vol.~34, no.~10, pp. 1993--2024, 2014.

\bibitem{bakas2017advancing}
S.~Bakas, H.~Akbari, A.~Sotiras, M.~Bilello, M.~Rozycki, J.~S. Kirby, J.~B.
  Freymann, K.~Farahani, and C.~Davatzikos, ``Advancing the cancer genome atlas
  glioma mri collections with expert segmentation labels and radiomic
  features,'' \emph{Scientific data}, vol.~4, no.~1, pp. 1--13, 2017.

\bibitem{bakas2018identifying}
S.~Bakas, M.~Reyes, A.~Jakab, S.~Bauer, M.~Rempfler, A.~Crimi, R.~T. Shinohara,
  C.~Berger, S.~M. Ha, M.~Rozycki \emph{et~al.}, ``Identifying the best machine
  learning algorithms for brain tumor segmentation, progression assessment, and
  overall survival prediction in the brats challenge,'' \emph{arXiv preprint
  arXiv:1811.02629}, 2018.

\bibitem{buda2019association}
M.~Buda, A.~Saha, and M.~A. Mazurowski, ``Association of genomic subtypes of
  lower-grade gliomas with shape features automatically extracted by a deep
  learning algorithm,'' \emph{Computers in Biology and Medicine}, vol. 109,
  2019.

\bibitem{prince1990constrained}
J.~L. Prince and A.~S. Willsky, ``Constrained sinogram restoration for
  limited-angle tomography,'' \emph{Optical Engineering}, vol.~29, no.~5, pp.
  535--544, 1990.

\end{thebibliography}

\appendices

\section{2D CT simulations with 120 view angles}

\begin{figure*}[t]
    \centering
    \includegraphics[width=0.9\linewidth]{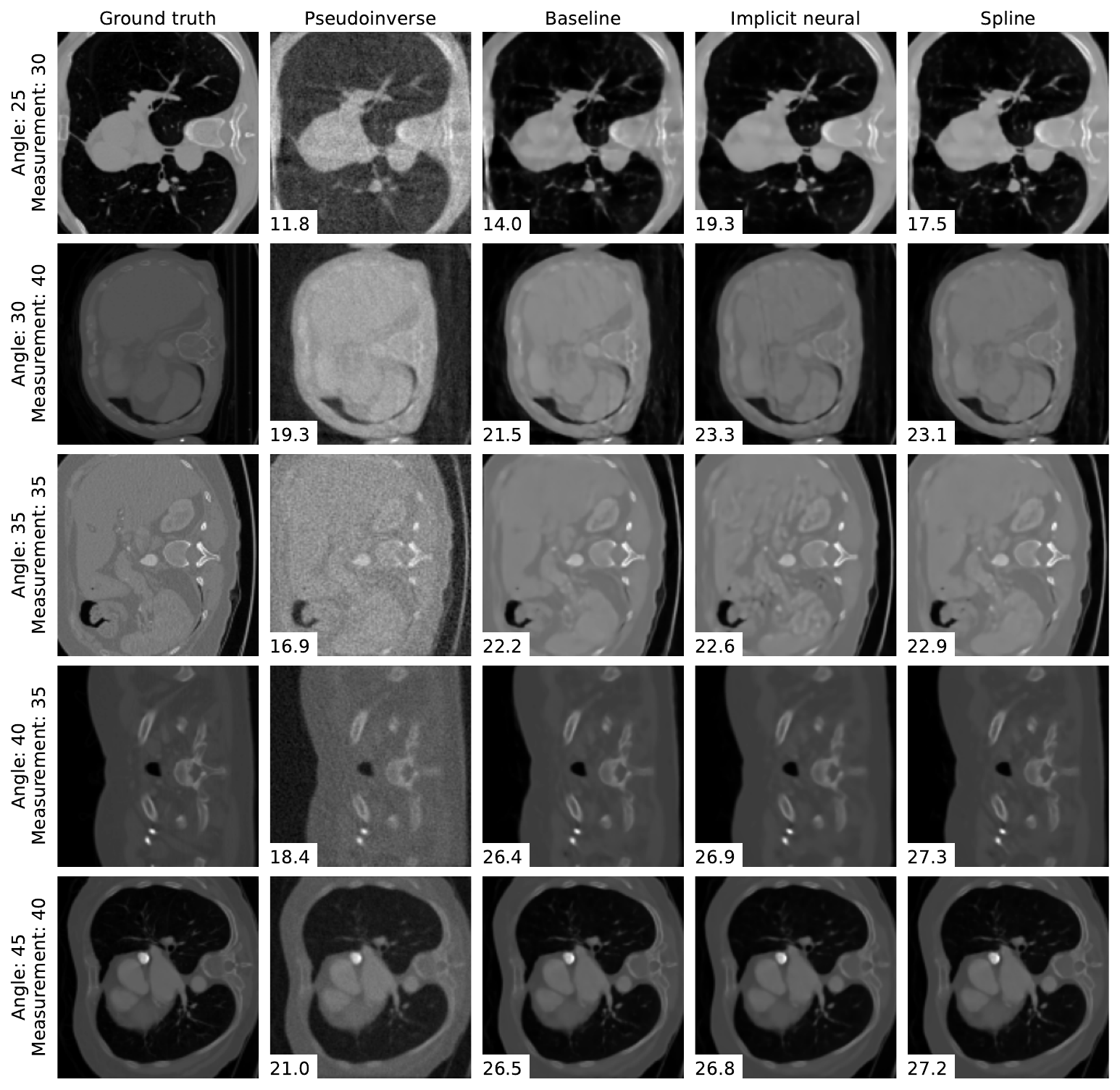}
    \caption{Example reconstructions for different measurement noise and \rev{2D CT view angle} uncertainty. \rev{There} are 120 view angles. The SNRs are shown for each reconstruction}
    \label{fig:image_sweep_recons_120}
\end{figure*}

In Fig. \ref{fig:image_sweep_recons_120} we show 2D CT imaging sample reconstructions for different combinations of measurement noise and measurement coordinate uncertainty when there are 120 view angles. The \rev{simulated} experiment is performed as described in Section \ref{sec:exp_xray_combos}. The reconstructions for when there are 90 view angles is shown in Fig.~\ref{fig:image_sweep_recons}.

\begin{figure}[t]
    \centering
    \includegraphics[width=0.9\linewidth]{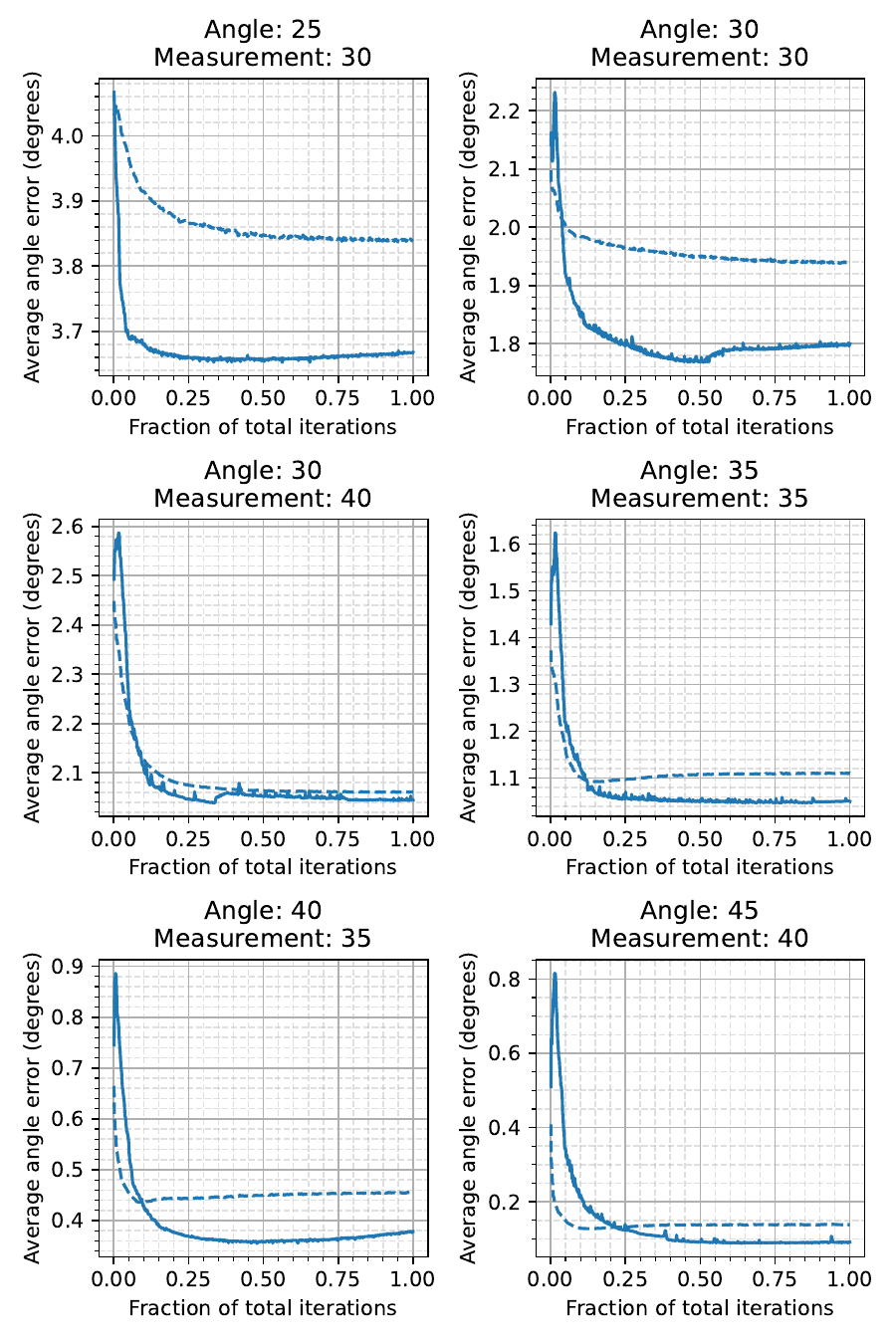}
    \caption{Average angle error for one test image with different measurement noise and measurement coordinate uncertainty combinations when there are 120 \rev{2D CT} view angles. The solid lines are for implicit neural representations and the dashed lines are for spline representations.}
    \label{fig:image_sweep_angle_error_120}
\end{figure}

Fig. \ref{fig:image_sweep_angle_error_120} shows average angle error plots for 2D CT imaging when there are 120 view angles. The experimental details are in Section \ref{sec:exp_xray_angle_error} and the equivalent results for 90 view angles are shown in Fig. \ref{fig:image_sweep_angle_error}.

\section{Three-dimensional CT imaging} \label{app:ct3d_experiments}

Similar to 2D CT imaging, in 3D CT imaging, two-dimensional projections of a three-dimensional volume are collected by tilting it at different tilt angles. The goal is to reconstruct the volume from the the series of projections. Inspired by CryoET imaging, we obtain projections at 60 tilt angles which we assume to be uniformly spaced on the interval $[-\nicefrac{\pi}{3}, \nicefrac{\pi}{3}]$. The true unknown tilt angles are perturbed from these. This setup is challenging because there is a `wedge' of tilt angles for which we do not have projections. The detectors are uniformly spaced on a two-dimensional unit square $[0, 1]^2$ and have no uncertainty \rev{in these experiments}. Combining these spaces gives the measurement coordinates space as $\Omega = [-\nicefrac{\pi}{3}, \nicefrac{\pi}{3}] \times [0,1]^2$.
\rev{The tilt angle uncertainty is simulated in the same way as the 2D CT imaging view angle uncertainty. Furthermore, we only optimize the tilt angles dimension of the measurement coordinates in \eqref{eq:joint_objective} because the detector locations have no uncertainty \rev{in these simulations}}.

To train the reconstruction 3D Unets, we use 3D volumes from the 2019 Brain Tumor Segmentation (BraTS) Challenge dataset \cite{menze2014multimodal, bakas2017advancing, bakas2018identifying}. The volumes are resized to be $64 \times 64 \times 64$. We use 478 volumes for training and 15 separate volumes to test and verify our framework.

\subsection{Combinations of measurement and coordinate error}

\begin{figure}
    \centering
    \includegraphics[width=1\linewidth]{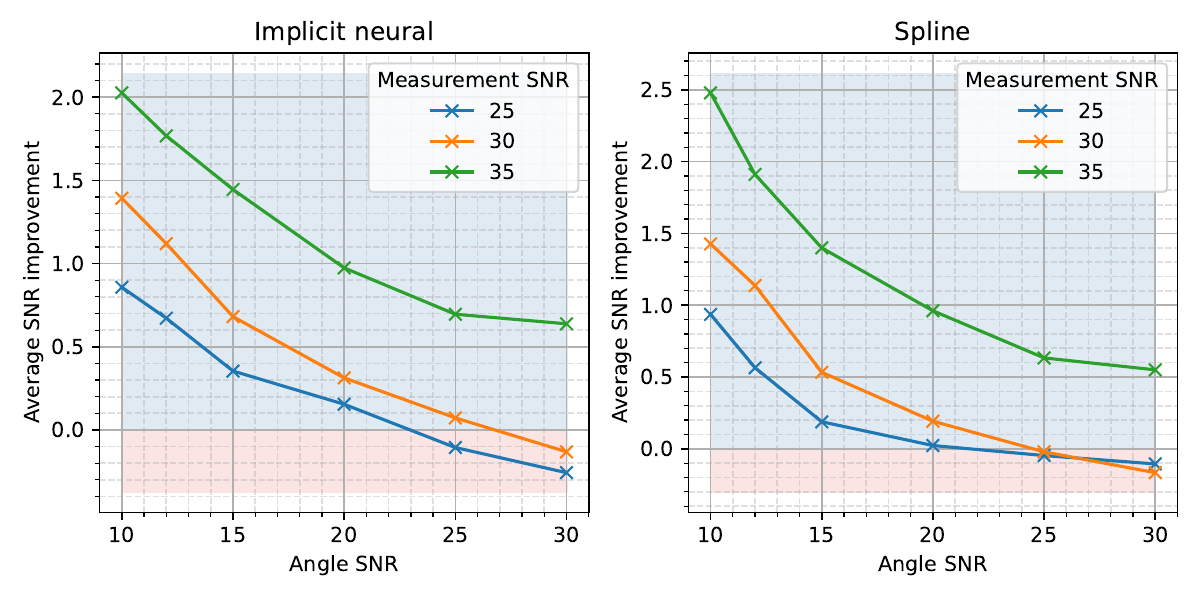}
    \caption{Average SNR improvement (dB) when using our framework~\eqref{eq:joint_objective} for \rev{3D CT. There are} different combinations of measurement noise SNR and measurement coordinate uncertainty SNR. There are 60 tilt angles on the interval $[-\nicefrac{\pi}{3}, \nicefrac{\pi}{3}]$.}
    \label{fig:image_sweep_3d}
\end{figure}

We consider different combinations of measurement noise and measurement coordinate uncertainty (in tilt angles) in the same way as was done for 2D CT. The trials are done over 15 test volumes and the results are shown in Fig. \ref{fig:image_sweep_3d}. We see that the performance trends for both implicit neural and spline representations are similar. Furthermore, the performance is consistent with what was seen for the 2D CT problem (Fig. \ref{fig:image_sweep})---as measurement coordinate uncertainty increases, our framework provides increasing gains.

\begin{figure*}[t]
    \centering
    \includegraphics[width=0.9\linewidth]{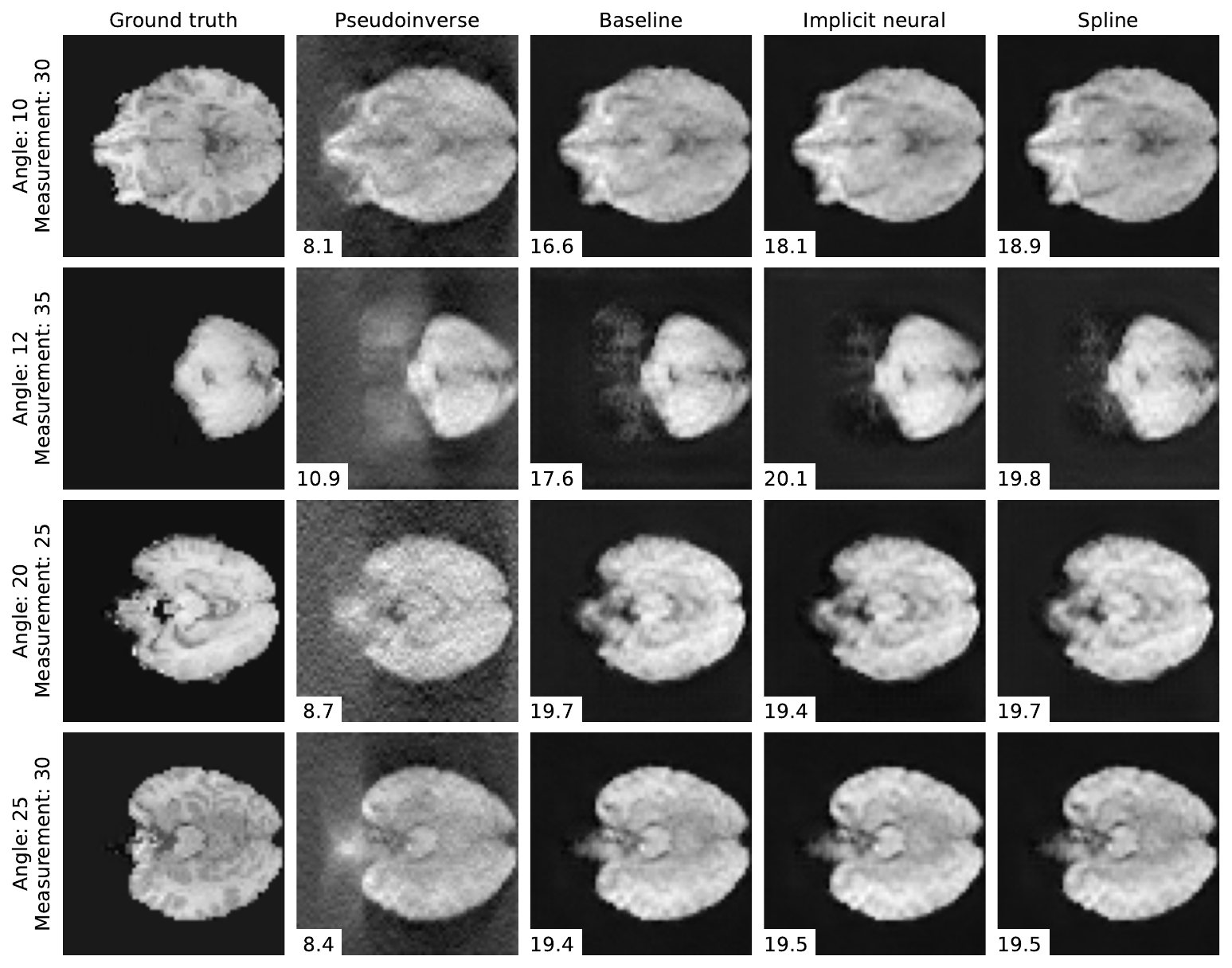}
    \caption{Example reconstructions of slices of the \rev{3D} test volumes. Reconstructions for different measurement noise SNR and measurement coordinate uncertainty SNR are shown. The reconstruction SNRs for each volume are stated.}
    \label{fig:image_sweep_recons_3d}
\end{figure*}

\begin{figure*}[t]
    \centering
    \includegraphics[width=0.8\linewidth]{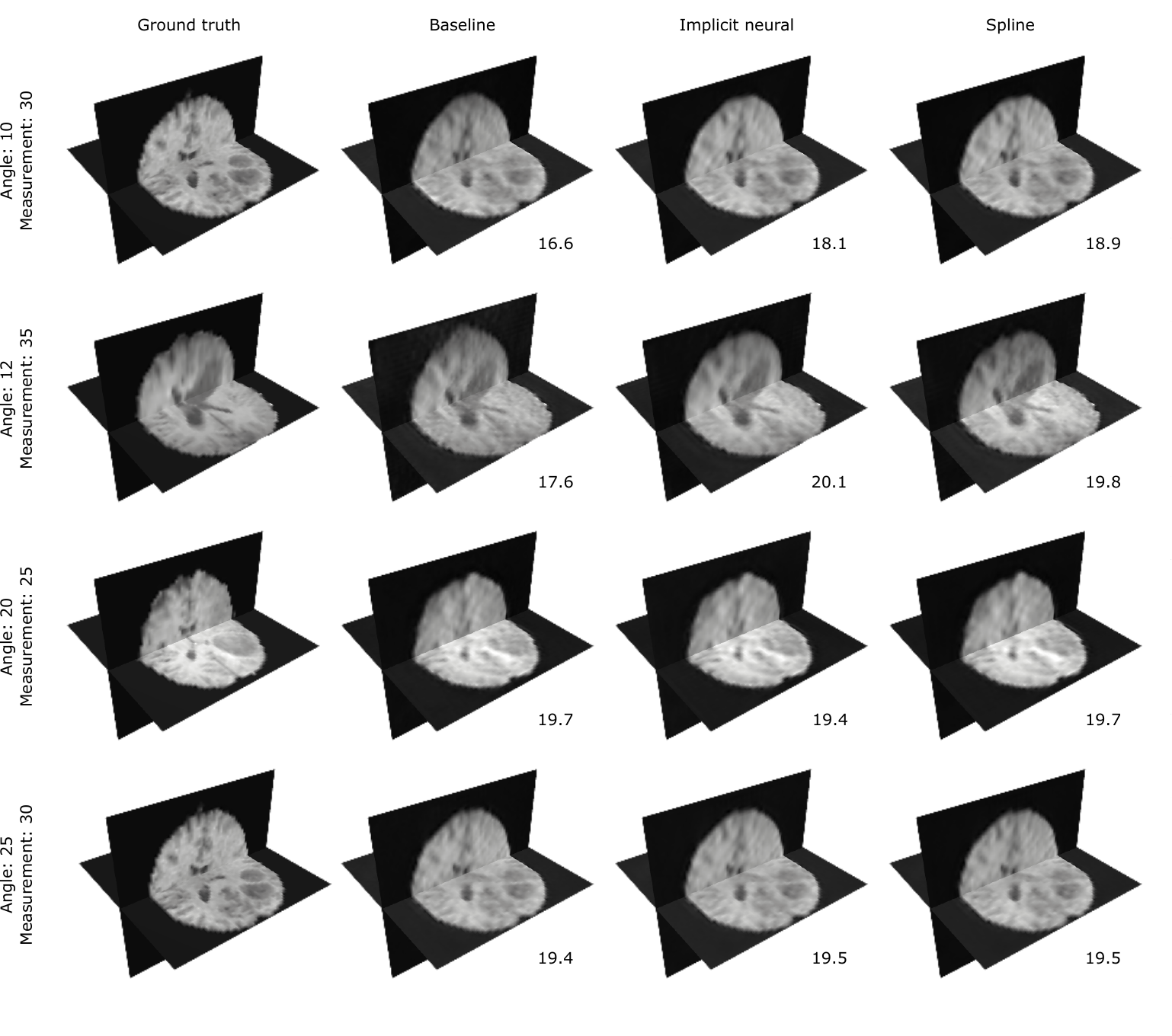}
    \caption{Example reconstructions for two of the orthogonal central slices of each of the \rev{3D} test volumes in Fig. \ref{fig:image_sweep_recons_3d} for \rev{3D} CT imaging. The SNRs for the entire volume are stated.}
    \label{fig:image_sweep_recons_3d_volumes_real}
\end{figure*}

Fig. \ref{fig:image_sweep_recons_3d} shows two-dimensional slices through some of the test volumes. The ground truth, psuedoinverse, baseline and our reconstructions are shown with the SNRs of the entire volumes. Similarly, Fig. \ref{fig:image_sweep_recons_3d_volumes_real} shows 3D reconstructions of the same test volumes. Two central orthogonal volume slices have been plotted. Reconstructions using our framework are better than the baseline at recovering the geometric structure of the volumes and have fewer artifacts.

\subsection{Learned tilt angles accuracy}

\begin{figure}[t]
    \centering
    \includegraphics[width=0.9\linewidth]{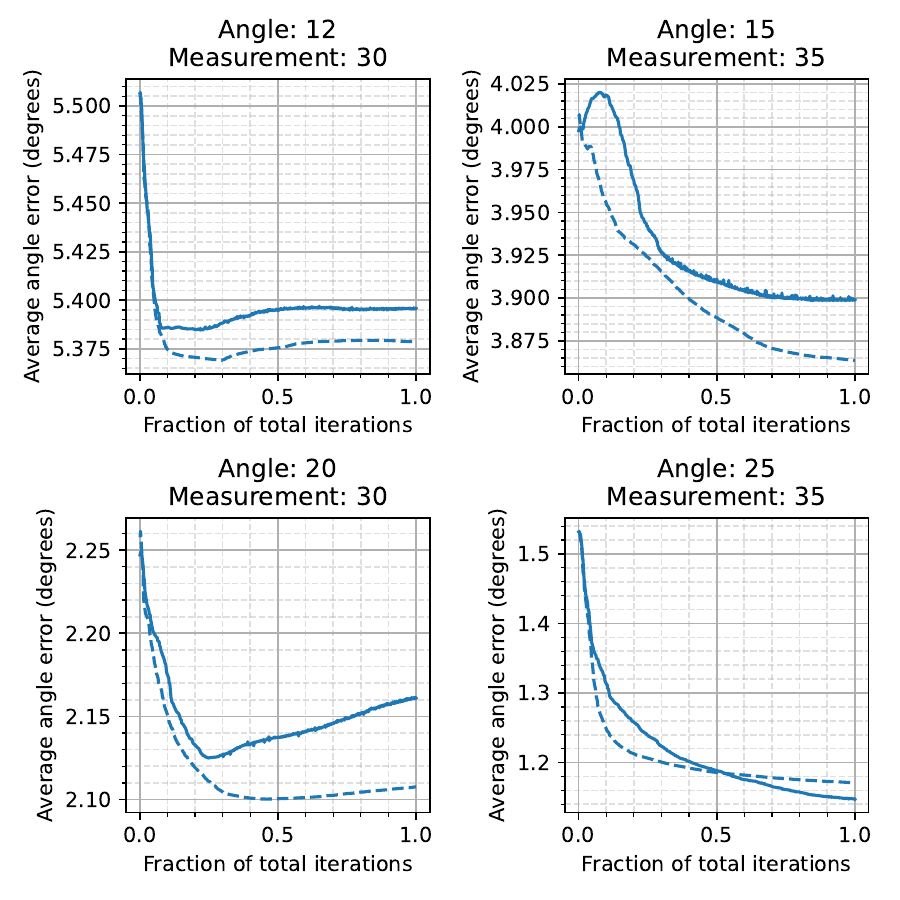}
    \caption{Average tilt angle error for one \rev{3D CT} test volume with different measurement noise and measurement coordinate uncertainty combinations. The solid lines are for implicit neural representations and the dashed lines are for spline representations.}
    \label{fig:image_sweep_angle_error_3d}
\end{figure}

Next we verify that the learned measurement coordinates are accurate. The uncertainty is only in the tilt angles and we follow the same procedure as when verifying the learned 2D CT view angles in Fig. \ref{fig:image_sweep_angle_error}. We use the metric defined in \eqref{eq:angle_error} and show the results in Fig. \ref{fig:image_sweep_angle_error_3d} for one test volume with different measurement noise and measurement coordinate uncertainty combinations. Again, the solid lines are for the implicit neural representations, and the dashed lines are for the spline representations. For both representation types, the average tilt angle error reduces as the optimization of \eqref{eq:joint_objective} progresses. This demonstrates that our framework learns tilt angles (measurement coordinates) that are more accurate than the assumed ones.

\section{Additional spline information} \label{app:splines}

We continue using 2D CT as an example and build on Section \ref{sec:splines} to provide further information on NURBS. Besides the learnable weights and learnable control point vectors explained in Section \ref{sec:splines}, there is a knot vector for each dimension of the measurement coordinates that is not learnable. The NURBS surface also has degrees $d_\Theta, d_\calT \in \Z^+$ for the view angle and detector location measurement coordinate dimensions. 
The knot vector for the view angle dimension has $(J + d_\Theta + 1)$ elements which are arranged in ascending order. We design its $u$th element, $k_u$, to be zero when $0 \leq u < d_\Theta + 1$, uniformly spaced between zero and one when $d_\Theta + 1 \leq u \leq J$, and one when $J < u \leq J + d_\Theta$ \cite{prasad2022nurbs}. The knot vector for the detector location dimension has $(J + d_\calT + 1)$ elements and is made in a similar manner by using its degree~$d_\calT$.

The rational basis functions defined in \eqref{eq:nurbs_basic} are
\begin{align}
    b_{j, k}([\theta', t']^\T)
    =
    \frac{
    w_{j, k} \, Q_{j, d_\Theta}(\theta') \, Q_{k, d_\calT}(t')
    }{
    \sum_{u=0}^{J - 1}
    \sum_{v=0}^{K - 1}
    w_{u, v}
    \, Q_{u, d_\Theta}(\theta') \, Q_{v, d_\calT}(t')
    } . \label{eq:rational_basis}
\end{align}
The function $Q_{u,d}(\cdot)$ is the $u$th B-spline basis function of degree $d$. If $k_u$ denotes the $u$th element of a knot vector, each basis function can be obtained using the Cox-de Boor recursion method,
\begin{align}
    Q_{u,0}(k) &= \begin{cases}
    1 & \text{if $k_u \leq k \leq k_{u + 1}$} \\
    0 & \text{otherwise}
    \end{cases} \label{eq:cdb_zero} \\
    Q_{u,d}(k) &= 
    \frac{k - k_u}{k_{u + d} - k_u} Q_{u,d-1}(k) \notag \\
    & \qquad +
    \frac{k_{u + d + 1} - k}{k_{u + d + 1} - k_{u + 1}} Q_{u+1,d-1}(k) . \label{eq:cdb_linear_interpolation}
\end{align}
From \eqref{eq:cdb_zero}, we can see that a degree zero NURBS is constructed from piecewise constant basis functions. The recursion \eqref{eq:cdb_linear_interpolation} is then used to create higher degree basis functions with larger support. If the denominator in any term of \eqref{eq:cdb_linear_interpolation} is zero, that term is taken to be zero.

The NURBS surface for two-dimensional measurement coordinates in \eqref{eq:nurbs_basic} is the tensor product of two one-dimensional NURBS curves as shown in \eqref{eq:rational_basis}. To create NURBS surfaces for $D$-dimensional measurement coordinates, we take the tensor product of $D$ one-dimensional NURBS curves. This is done for $D=3$ in Appendix~\ref{app:ct3d_experiments}.

\section{Implementation details} \label{app:parameters}

\subsection{Framework optimization}

In this section we provide implementation details for solving our optimization problem \eqref{eq:joint_objective}. All parameters were tuned on a held out set of images for three randomly chosen measurement and operator error combinations. 

To implement NURBS, we modified and extended the PyTorch source code released by the NURBS-Diff module authors \cite{prasad2022nurbs}. The implicit neural representation and neural network for $G_{\vmu}(\cdot)$ in \eqref{eq:joint_objective} are also implemented in PyTorch. This enables us to conveniently optimize the objective function \eqref{eq:joint_objective} using automatic differentiation and the Adam optimizer.

When implicit neural networks are used, the optimization is run for at least 8,000 iterations and at most 20,000 iterations. The optimization is terminated when the loss value between successive iterations is below $1 \times 10^{-10}$ for 2D CT imaging and $1 \times 10^{-11}$ for 3D CT imaging. When splines are used \rev{and there is any view angle or tilt angle uncertainty,} the optimization runs for at least 2,000 iterations and at most 5,000 iterations. \rev{If there is only detector location uncertainty, the optimization runs for at least 6,000 iterations and at most 15,000 iterations. Additionally, when splines are used, the} optimization for all imaging problems is terminated when the loss value between successive iterations is below $1 \times 10^{-11}$.

\subsubsection{2D CT imaging} \label{app:ct_details}

When implicit neural representations are used, \rev{$\lambda = 0.1$} in \eqref{eq:joint_objective}. The learning rate for both the neural network parameters and the input coordinates is $5 \times 10^{-4}$.
When splines are used, \rev{$\lambda = 0.025$ if there is only view angle uncertainty, and $\lambda = 0.25$ if there is any detector location uncertainty.} The learning rate for the neural network parameters is $5 \times 10^{-2}$ and for the input coordinates is $2 \times 10^{-4}$.

For the implicit neural representation, we use the cosine of the view angle rather than the angle when creating the measurement coordinate. This encodes the circular nature of angular data and improves performance.

\subsubsection{3D CT imaging}

When implicit neural representations are used, \rev{$\lambda = 0.1$}. The learning rate for the neural network parameters is $1 \times 10^{-3}$ and for the input coordinates is $1 \times 10^{-4}$.
When splines are used, \rev{$\lambda = 0.6$}. The learning rate for the neural network parameters is $5 \times 10^{-3}$ and for the input coordinates is $1 \times 10^{-4}$.

\subsection{Reconstruction Unet training}

In this section we describe the implementation details for the Unet neural networks used for the reconstruction method $G_{\vmu}(\cdot)$ in \eqref{eq:joint_objective}.

A mean-squared error loss function is minimized using Adam during training. We train the Unets for 100 epochs where one epoch is a full pass through the training dataset.

For the 2D Unet, a batch size of 128 and learning rate of $1 \times 10^{-3}$ is used. For the 3D Unet, a batch size of 16 and learning rate of $1 \times 10^{-3}$ is used.

Publicly available Unet architectures were downloaded and trained. Unless mentioned here, the default parameters from the download sources were used. The 2D Unet model is from \url{https://github.com/mateuszbuda/brain-segmentation-pytorch} \cite{buda2019association}. We used one input channel, one output channel, and 16 features in the first layer. The 3D Unet model is from \url{https://github.com/ELEKTRONN/elektronn3}. We used one input channel, one output channel, 16 features in the first layer and a depth of four blocks.

Due to the limited size of the 3D volume training dataset for 3D CT imaging, we use a data augmentation strategy. We perform random horizontal flips, random vertical flips, and random rotations by 90, 180 or 270 degrees.

\subsection{NURBS}

\subsubsection{2D CT imaging}

For 2D CT imaging, the NURBS degree along the \rev{measurement coordinate dimension being learned} is 18. It is two in the \rev{measurement coordinate dimension that is not being learned}. Furthermore, \rev{where there is view angle uncertainty}, we create additional control points to ensure the spline measurement representations satisfy the Radon transform measurement consistency conditions \cite{prince1990constrained}
\begin{align}
    r_{\vvarphi}([\theta + \pi, t]^\T) &=  r_{\vvarphi}([\theta, -t]^\T) , 
\end{align}
and 
\begin{align}
    r_{\vvarphi}([\theta - \pi, t]^\T) &= r_{\vvarphi}([\theta, -t]^\T) ,
\end{align}
where a detector location of $-t$ refers to the $t$th last detector. This gives the locally supported NURBS basis functions \eqref{eq:rational_basis} a sufficient number of control points around $0$ and $\pi$ radians and ensures the NURBS is accurate in the interval $[0, \pi]$ radians. Specifically, if $d_\Theta$ is the degree of the spline in the view angle dimension, we use the consistency condition to create new control points for \eqref{eq:nurbs_basic}
\begin{align}
    \vp_{j+J-1,k} = \left[\theta_j + \pi, t_k, \vy([\wt{\theta}_j, -\wt{t}_k]^\T)\right]^\T
\end{align}
for $1 \leq j \leq d_\Theta$ and
\begin{align}
    \vp_{j - J-1,k} = \left[\theta_j - \pi, t_k, \vy([\wt{\theta}_j, -\wt{t}_k]^\T)\right]^\T
\end{align}
for $J - d_\Theta < j \leq J$ where $J$ is the number of view angles in the observed measurements. Essentially, if there are $K$ detectors, we create $2 d_\Theta K$ additional control points.

\subsubsection{3D CT imaging}

For 3D CT, the NURBS degree along the tilt angle measurement coordinate dimension is 15. It is two in the two detector location dimensions. Additionally, for this imaging technique, we fix the basis function weights to one and do not optimize them. This makes the NURBS surface a B-spline surface. 

\end{document}